\documentclass[
    reprint,
    pra,
    aps,
    notitlepage,
    nofootinbib
]{revtex4-2}

\usepackage{amsmath}
\usepackage{amssymb}
\usepackage{bm,braket}
\usepackage{float} 
\usepackage{tikz} 
\usepackage{multirow}
\usetikzlibrary{quantikz}

\newcommand{\figref}[1]{Figure~#1}
\newcommand{\secref}[1]{Section~#1}
\newcommand{\appref}[1]{Appendix~#1}

\newcommand{\subfigimg}[3][,]{%
  \setbox1=\hbox{\includegraphics[#1]{#3}}
  \leavevmode\rlap{\usebox1}
  \rlap{\hspace*{0pt}\raisebox{\dimexpr\ht1-2\baselineskip + 12pt}{#2}}
  \phantom{\usebox1}
}
\newcommand{\lsubfigimg}[3][,]{%
  \setbox1=\hbox{\includegraphics[#1]{#3}}
  \leavevmode\rlap{\usebox1}
  \rlap{\hspace*{-10pt}\raisebox{\dimexpr\ht1-2\baselineskip + 12pt}{#2}}
  \phantom{\usebox1}
}

\usepackage{hyperref}
\usepackage{multirow}
\usepackage{enumitem}
\hypersetup{
     colorlinks   = true,
     citecolor    = blue
}
\def\be{\begin{equation}}
\def\ee{\end{equation}}
\def\bea{\begin{eqnarray}}
\def\eea{\end{eqnarray}}
\def\bes{\begin{eqnarray}}
\def\ees{\end{eqnarray}}
\def\bi{\begin{itemize}}
	\def\ei{\end{itemize}} 
 
\usepackage{amsthm}

\theoremstyle{definition}

\DeclareMathOperator{\Tr}{Tr}

\newcommand{\norm}[1]{\left\lVert #1 \right\rVert}

\usepackage{tikz}
\usetikzlibrary{braids}

\renewcommand\Re{\operatorname{Re}}
\renewcommand\Im{\operatorname{Im}}

\definecolor{softblue}{rgb}{0.85, 0.90, 0.98}
\definecolor{softgray}{rgb}{0.93, 0.93, 0.93}
\definecolor{softred}{rgb}{0.97, 0.81, 0.8}
\definecolor{softgreen}{rgb}{0.83, 0.91, 0.83}
\definecolor{softpurple}{rgb}{0.88, 0.83, 0.90}
\definecolor{softorange}{rgb}{1.0, 0.90, 0.8}

\begin{document}
\title{Efficient Simulation of Open Quantum Systems on NISQ Trapped-Ion Hardware}

\author{Colin Burdine}
\email{colin\_burdine1@baylor.edu}
\affiliation{%
 Department of Electrical and Computer Engineering, Baylor University, Waco, Texas 76798-7316, USA
}%

\author{Nora Bauer}
\affiliation{Department of Physics and Astronomy,  The University of Tennessee, Knoxville, TN 37996-1200, USA}

\author{George Siopsis }
\affiliation{Department of Physics and Astronomy,  The University of Tennessee, Knoxville, TN 37996-1200, USA}

\author{Enrique P. Blair}%
\affiliation{%
 Department of Electrical and Computer Engineering, Baylor University, Waco, Texas 76798-7316, USA
}%

\begin{abstract}
    Simulating open quantum systems, which interact with external environments, presents significant challenges on noisy intermediate-scale quantum (NISQ) devices due to limited qubit resources and noise. In this paper, we propose an efficient framework for simulating open quantum systems on NISQ hardware by leveraging a time-perturbative Kraus operator representation of the system's dynamics. Our approach avoids the computationally expensive Trotterization method and exploits the Lindblad master equation to represent time evolution in a compact form, particularly for systems satisfying specific commutation relations. We demonstrate the efficiency of our method by simulating quantum channels, such as the continuous-time Pauli channel and damped harmonic oscillators, on NISQ trapped-ion hardware, including IonQ Harmony and Quantinuum H1-1. Additionally, we introduce hardware-agnostic error mitigation techniques, including Pauli channel fitting and quantum depolarizing channel inversion, to enhance the fidelity of quantum simulations. Our results show strong agreement between the simulations on real quantum hardware and exact solutions, highlighting the potential of Kraus-based methods for scalable and accurate simulation of open quantum systems on NISQ devices. This framework opens pathways for simulating more complex systems under realistic conditions in the near term. 
\end{abstract}
\maketitle 

\maketitle

\section{Introduction}
\label{sec:introduction}

The efficient simulation of open quantum systems is crucial for the study of quantum algorithms, materials, communications, and many far-reaching aspects of quantum technology.  
This area is still developing, especially given the challenges associated with noisy intermediate-scale quantum (NISQ) devices. Simulating open quantum systems—which involve interactions with an external environment—presents additional complexity compared to closed quantum systems, making this an important and active area of research.

Open quantum systems with Markovian (memoryless) noise are typically described by the Lindblad master equation, which models the system's evolution in the presence of interaction with the environment. Work on simulating Lindblad-type dynamics using quantum simulators and NISQ hardware has involved studies of quantum noise channels, such as depolarizing, amplitude damping, and phase damping channels \cite{cleve_et_al:LIPIcs.ICALP.2017.17}. These studies are foundational because noise and decoherence naturally occur in quantum hardware, making it essential to simulate and understand such effects.

A popular approach to simulating open quantum systems on quantum devices uses Trotterization (as in closed systems) to discretize the time evolution of the system. While successful to some extent, this method becomes impractical on NISQ devices due to the large number of gates required \cite{Han_2021,Childs_2021}. Other methods that have been explored are based on variational quantum algorithms which can be designed to be hardware-efficient, though these methods can still struggle with noise and require classical optimization \cite{Chen_2024,Endo_2020,Mellak2024DeepNN}. An alternative approach is based on quantum trajectory (Monte Carlo) methods \cite{Carmichael_1993}, which simulate the evolution of a quantum system under stochastic noise or decoherence. These methods have been demonstrated on small-scale quantum devices, though their practical application is still limited by device noise \cite{Cech_2023}. Other methods include using the NISQ device noise for probabilistic error cancellation \cite{Guimaraes_2023}, matrix product state (MPS) numerical methods \cite{Cygorek2021SimulationOO}, and quantum imaginary time evolution (QITE) \cite{Kamakari_2022}. 

Certain Trotterless schemes have been devised that should be in principle better suited to NISQ hardware \cite{Hu2019AQA}, but their application is limited to toy models, lacking generality. Simulation schemes for analog quantum simulators, such as trapped ions and Rydberg atoms, have been investigated under various environmental conditions \cite{Kim_2022}. These platforms allow for the exploration of dissipative processes and quantum systems interacting with external fields, though their scalability and programmability are limited compared to digital quantum devices.

Various challenges exist. Simulating open quantum systems often requires complex quantum circuits, which are sensitive to noise on NISQ devices, limiting the depth and number of qubits that can be effectively used. Developing error mitigation techniques is a central focus for making such simulations practical on NISQ devices. Finding efficient representations for dissipation processes (e.g., using Kraus operators or other reduced models) has been a significant research focus. While various studies have simulated specific types of open quantum systems on NISQ hardware, the methods often rely on approximations, and large-scale, high-fidelity simulations remain a challenge. 

In this work, we introduce more efficient methods for simulating open quantum systems, including novel circuit designs and error mitigation strategies. These approaches aim to push the boundaries of what is possible on NISQ devices and overcome some of the limitations of earlier methods. We focus on Kraus representation-based methods to simulate the dynamics of these systems. These methods exploit a representation of the system's time evolution in terms of Kraus operators, which model the interaction between the system and its environment. When specific commutation relations between the system's Hamiltonian and Lindblad operators are satisfied, this representation can be computed efficiently without Trotterization. The Kraus operators offer a more efficient and NISQ-friendly way of simulating these systems. We apply techniques to map the Kraus operators and time evolution operators to quantum circuits for implementation on actual NISQ hardware (IonQ Harmony and  Quantinuum H1-1 devices).

Our results confirm that the Kraus representation-based method, combined with error mitigation, can effectively simulate open quantum systems on NISQ devices, with strong agreement between experimental results and theoretical predictions. They represent a significant step forward in the simulation of open quantum systems on NISQ devices and a critical step in bridging theoretical methods and real-world quantum computing applications.

This paper is organized as follows: In \secref{\ref{sec:background}}, we give a review of the motivating background theory, including the Lindblad equation, Trotterization, and Kraus representations.  We also discuss how systems satisfying certain commutation relations admit Kraus representations that can be simulated efficiently, and we discuss some important special cases of these systems. In \secref{\ref{sec:results}}, we present the results of applying our framework to simulate selected open quantum systems on real quantum hardware. We simulate a two-qubit system undergoing decoherence, modeled as a Pauli channel. The results show good agreement between the Kraus series-based simulation and the exact Lindblad equation solution. We also simulate a 2D quantum harmonic oscillator under damping, which is modeled using angular momentum operators as Lindblad operators. The results are compared with an exact solution, and strong agreement is observed, even in the presence of noise. For larger systems, the effect of noise becomes more significant.
We present hardware-agnostic error mitigation techniques, such as Pauli channel fitting and quantum depolarizing channel inversion, that can be applied to improve simulation fidelity. Finally, in \secref{\ref{sec:conclusion}}, we summarize our results and give directions for future research. Appendix \ref{sec:appendix_circuits} contains additional details for mapping Kraus operators to quantum circuits for the relevant systems. Appendix \ref{sec:appendix_pauli_noise_models} contains additional details for applying Pauli noise models to density matrix simulations. 

\section{Background}
\label{sec:background}

\subsection{Lindblad Equation}
\label{sec:background_lindblad_equation}
In this paper, we will consider open quantum systems under the Born-Markov approximation-- that is, systems weakly coupled to a large environment in which correlations between the system and environment are short-lived. Under these assumptions, we are free to impose that the the dynamics of a system are Markovian (i.e. the environment is effectively ''memoryless'') and can be modeled by a completely-positive trace preserving (CPTP) map. In non-relativistic systems governed by the Schrodinger equation, the CPTP time evolution of a system can always be written in the Lindblad form
\begin{equation}
    \begin{aligned}
    \dot{\rho}(t) &= -\frac{i}{\hbar}[H, \rho(t)]  \\
    &\qquad + \sum_{n=1}^{N} \gamma_n \left( L_n \rho(t)L_n^\dagger - \frac{1}{2}\lbrace L_n^\dagger L_n, \rho(t) \rbrace \right),
    \end{aligned}
    \label{eqn:lindblad}
\end{equation}
where $\rho(t)$ is the system density matrix, $H$ is the system Hamiltonian, and the $L_n$ are Lindblad operators with positive damping coefficients $\gamma_n$. The Lindblad operators are often imposed phenomenologically on a system to model environment-induced decoherence \cite{breuer_theory_2002}.

\subsection{Trotter-Based Methods}
\label{sec:background_trotter_based_methods}

Solving Eq. \eqref{eqn:lindblad} for arbitrary quantum systems as a closed-form function of time is known to be a difficult problem. As a result, one must commonly resort to numerical integration methods on quantum or classical computers. On quantum computers, the most popular of these methods are those that employ Trotter product formulas \cite{trotter_product_1959}.
This is usually achieved by re-writing the right-hand side of Eq. \eqref{eqn:lindblad} in the equivalent superoperator form, namely
\begin{equation}
    \dot{\vec{\rho}}(t) = \mathcal{D}\vec{\rho}(t),
\end{equation}
where $\mathcal{D}$ is a superoperator matrix that acts linearly on the ``vectorized" density matrix $\vec{\rho} = \sum_{i,j} \rho_{ij} (\ket{i} \otimes \ket{j})$. Next, the superoperator $\mathcal{D}$ is decomposed into a sum of non-commuting operators, e.g.
\begin{equation}
    \mathcal{D} = \mathcal{D}_1 + \mathcal{D}_2
    \label{eqn:D_decomp}
\end{equation}
such that the evolution of each component $\mathcal{D}_i$ for a small time step $\delta t$ can be realized as an efficient sequence of quantum gates $U_{\mathcal{D}_i} \simeq \exp(\delta t \mathcal{D}_i)$. The key idea of a Trotter product formula is that one can repeatedly apply an alternating sequence of the gates $U_{\mathcal{D}_i}$ over many discrete $\delta t$ time steps such that the approximation error vanishes in the limit as $\delta t \rightarrow 0$. For example, applying a second order Trotter product formula \cite{hatano_finding_2005} to Eq. \eqref{eqn:D_decomp} over an $n$-step time evolution period $t$ ($\delta t = t/n$), we obtain the approximation:
\begin{equation}
    \vec{\rho}(t) = \left[e^{(\delta t/2) \mathcal{D}_1}e^{\delta t \mathcal{D}_2}e^{(\delta t/2) \mathcal{D}_1}\right]^n\vec{\rho}(0) + O(t(\delta t)^2)
\end{equation}
This approach to simulating open quantum system is appealing in that it only requires the repeated application of the quantum gate unitaries $U_{\mathcal{D}_i}$. However, it can be quite costly for long evolution periods $t$ or systems where the unitaries $U_{\mathcal{D}_i}$ contain many gates \cite{childs_theory_2021}. This is especially known to be the case for dense bosonic systems, where computing the exponential of bosonic creation and annihilation operator terms in $\mathcal{D}$ is known to be quite expensive \cite{sawaya_resource-efficient_2020}. For this reason, Trotter-based techniques have remained difficult to apply on NISQ devices, except in the cases where the system dynamics can be straightforwardly mapped to device-native operations that can be performed with minimal noise.

\subsection{Kraus Representation-Based Methods}
\label{sec:background_kraus_methods}

For some systems, Trotterization can produce quantum circuits that require a large number of gates and qubits, which is undesirable for NISQ applications. However, for some special classes of systems, a closed-form solution to the Lindblad equation can be found that significantly reduces the complexity of Trotterization or even avoids Trotterization altogether. This closed form solution can always be written in the form of a time-dependent Kraus operator representation of the solution to Eq. \eqref{eqn:lindblad}, given by:
\begin{equation}
    \rho(t) = \sum_i K_i(t) \rho(0) K_i(t)^{\dagger}
    \label{eqn:kraus_operator_sum}
\end{equation}
where the $K_i(t)$ are time-dependent Kraus operators that satisfy the unital condition $\sum_{i} K_i(t)K_i(t)^{\dagger} = I$.

As a direct consequence of Choi's theorem \cite{choi_completely_1975}, one can always solve for a minimal set of $d^2$ Kraus operators $K_i$ corresponding to the time evolution of a $d \times d$ density matrix for a period $t$; however, finding this minimal set of Kraus operators for arbitrary systems is classically hard, requiring the diagonalization of the system's $d^2 \times d^2$ Choi matrix for each time step $t$ \cite{havel_robust_2003}. Fortunately, it can be shown that when a system's Lindblad operators and Hamiltonian satisfy certain commutation relations, a series of non-minimal Kraus operators corresponding to the time-perturbative treatment of the system's time evolution can be solved for exactly, without requiring the use of diagonalization. Specifically, if a Hamiltonian and a set of Lindblad operators satisfy the commutation relations
\begin{enumerate}[label=(\roman*)]
    \item $[H,L_n^\dagger L_n] = 0$ (for all $L_n$)
    \item $[L_n^\dagger L_n, L_{n'}^\dagger L_{n'}] = 0$\quad (for all $L_n, L_{n'}$)
    \item $[H, L_n] = \nu L_n$\\[2mm] 
        (for some $\nu \in \mathbb{C}$ with $\Im(\nu) \ge 0$, for all $L_n$)
    \item $\sum_{n'} \gamma_{n'}[L_{n'}^\dagger L_{n'}, L_n] = \lambda L_n$\\[2mm] 
        (for some $\lambda \in \mathbb{C}$ with $\Re(\lambda) \le 0$, for all $L_n$)
\end{enumerate}
then Eq. \eqref{eqn:lindblad} can be solved in closed form by a time-perturbative Kraus series of the general form 
\begin{equation}
    \rho(t) = \sum_{m=0}^{\infty}\sum_{\vec{k} \in \{1,2,...,N\}^m} K_{m,\vec{k}}(t)\rho(0) K_{m,\vec{k}}(t)^{\dagger}
    \label{eqn:tp_kraus_series}
\end{equation}
where $m$ is the order of each term and $\vec{k}$ is an index vector corresponding to each product sequence of $m$ Lindblad operators. In the general case, each Kraus operator term in the time-perturbative series takes the form
\begin{equation}
    K_{m,\vec{k}}(t) = T(t) \sqrt{\frac{f(t)^m}{m!}} \prod_{i=1}^m \left( \sqrt{\gamma}_{\vec{k}_i} L_{\vec{k}_i} \right).
    \label{eqn:tp_kraus_operator}
\end{equation}
Above, $f(t)$ is a positive function of time and $T(t)$ is the effective Hamiltonian evolution operator given by
\begin{equation}
    T(t) = \exp\left(-\frac{it}{\hbar}H_{\text{eff}}\right)
    \label{eqn:T_evolution_operator}
\end{equation}
where the effective Hamiltonian $H_{\text{eff}}$ is
\begin{equation}
    H_{\text{eff}} = H - \frac{i\hbar}{2}\sum_{n=1}^N \gamma_n L_n^{\dagger}L_n.
    \label{eqn:H_eff}
\end{equation} 
Because conditions (i) and (ii) are satisfied, the operator $H_{\text{eff}}$ is diagonalizable in the same basis as the original Hamiltonian $H$ with eigenvalues of the form 
\begin{equation}
    E_{\text{eff}} = E - \frac{i\hbar}{2}\sum_{n=1}^N \gamma_n\xi_n 
    \label{eqn:E_eff_eigenvalues}
\end{equation} 
where $E$ is a real eigenvalue of $H$ and each $\xi_n$ is a non-negative real eigenvalue of $L_n^{\dagger}L_n$. Since the imaginary component of each eigenvalue $E_{\text{eff}}$ is non-positive, the effective Hamiltonian evolution operator $T(t)$ is a contraction, meaning $\norm{T(t) \rho T(t)^{\dagger}} \leq \norm{\rho}$ for any density matrix $\rho$ and time $t \geq 0$. The time-dependence of the Kraus operator terms in Eq. \eqref{eqn:tp_kraus_operator} is described by the positive function $f(t)$, which takes different forms, depending on the values of $\nu$ and $\lambda$ as defined in conditions (iii) and (iv). In the most general case, $f(x)$ takes the form
\begin{equation}
    f(t) = \begin{cases}
        t & \alpha = 0 \\
        (1-e^{-\alpha t})/\alpha & \alpha > 0
    \end{cases}
    \label{eqn:kraus_series_f}
\end{equation}
where $\alpha = 2\Im(\nu)/\hbar - \Re(\lambda)$ is a positive constant. When $\alpha = 0$, we observe that the worst case time-dependence of the Kraus operators is such that $K_{m,\vec{k}}(t) \sim \sqrt{t^{m}/m!}$. This case corresponds to systems with unitary symmetries (e.g. Pauli channels), where the unitary symmetry operators are used as Lindblad operators with different choice of damping parameters $\gamma_n$. The $\alpha > 0$ case corresponds to systems where $H$ and the Lindblad terms $L_n^{\dagger}L_n$ exhibit non-trivial commutation relations (e.g. bosonic systems with finite particle number). For these systems, the time dependence can be reduced to $K_{m,\vec{k}}(t) \sim \sqrt{1/m!}$, which means that the Kraus series converges much faster and exhibits much lower truncation error than the general case. For additional details regarding the derivation of the general form of the time-perturbative Kraus operators in Eq. \eqref{eqn:tp_kraus_operator}, we refer the reader to reference \cite{burdine_trotterless_2024}.

\subsubsection{Mapping to Quantum Circuits}
\label{sec:background_mapping_to_circuits}

Upon inspection of Eq. \eqref{eqn:tp_kraus_series} and Eq. \eqref{eqn:tp_kraus_operator}, we observe that if the effective time evolution operator $T(t)$ and the Lindblad operators $L_n$ can be efficiently realized as quantum circuits acting on a pure state $\ket{\psi}$ (or a purified superposition $\sum_i \sqrt{p_i} \ket{\psi_i}$), then one can straightforwardly simulate the time-evolution of an arbitrary density matrix $\rho = \sum_{i} p_i \ket{\psi_i}\bra{\psi_i}$ on a quantum computer. This is achieved by first applying the sequence of Lindblad operators $\prod_{i} L_{\vec{k}_i}$ followed by $T(t)$, and then multiplying by the scalar term 
\begin{equation}
a_{m,\vec{k}} = \sqrt{\frac{f(t)^m}{m!}} \prod_{i = 1}^m \sqrt{\gamma_{\vec{k}_i}}.
\label{eqn:a_mk_scalar}
\end{equation}
This must be repeated for each Kraus term $K_{m,\vec{k}}(t)$ up to a desired cutoff order $m=M$. Applying each of these operators through quantum gates, however, is a non-trivial task and requires additional discussion.

First, we consider how the Lindblad operators $L_n$ can be expanded into unitary operations. Since it may generally be the case that $\norm{L_n} \neq 1$, the Lindblad operators must first be normalized using the transformation
\begin{equation}
\begin{aligned}
    L_n &\mapsto L_n/b_n \\
    \gamma_n &\mapsto b_n^2 \gamma_n
\end{aligned}
\label{eqn:lindblad_rescaling}
\end{equation}
where $b_n = \norm{L_n}$ is the normalization constant. This transformation has no effect on the dynamics of Eq. \eqref{eqn:lindblad}, and ensures that each $L_n$ can be encoded in a unitary of norm $1$. Once the $L_n$ operators are normalized, they must be expanded to unitaries that can be implemented with quantum gates. This is most commonly achieved through a unitary block-encoding scheme in which the transformed Lindblad operators are embedded in a block unitary $U_{L_n}$ that requires at least one clean ancilla qubit per application in a quantum circuit \cite{camps_fable_2022}. One such block form is given by the unitary Sz.-Nagy dilation \cite{schaffer_unitary_1955} of $L_n$:
\begin{equation}
    U_{L_n} = \begin{pmatrix}
        L_n & \sqrt{I - L_nL_n^{\dagger}} \\
        \sqrt{I - L_n^{\dagger}L_n} & -L_n^{\dagger}
    \end{pmatrix}.
    \label{eqn:sznagy_dilation}
\end{equation}

This block encoding is such that $U_{L_n}(\ket{0} \otimes \ket{\psi}) = (L_n\ket{\psi}) \otimes \ket{0} + \ket{\phi} \otimes \ket{1}$, where $\ket{\phi} = \sqrt{I - L_n^{\dagger}L_n}\ket{\psi}$ is a state that must be discarded when the ancilla qubit measures $\ket{1}$. Each unitary $U_{L_n}$ can be decomposed efficiently as sequences of quantum gates using algorithms such as the linear combination of unitaries (LCU) method \cite{zheng_universal_2021, childs_hamiltonian_2012} or the quantum Shannon decomposition (QSD) \cite{shende_synthesis_2006}. Due to the constraints of NISQ quantum hardware, implementing arbitrary unitary operators without accumulating noise is a significant challenge; thus the implementation of complex Lindblad operators (for example the bosonic annihilation operator $\hat{a}$) requires careful circuit design and hardware-specific optimization \cite{sawaya_resource-efficient_2020}. We give additional discussion regarding circuit design in \appref{\ref{sec:appendix_circuits}}.

Next, we briefly discuss how the operator $T(t)$ can be implemented. $T(t)$ is a contraction, meaning it can be expanded to a unitary operator without normalization. This expansion can be achieved through a unitary block-encoding scheme, however we recall that the effective Hamiltonian $H_{\text{eff}}$ is diagonalizable in the same basis as $H$. As a result, if a diagonal form of $H$ is known, then a diagonal block-encoding of the the operator $T(t)$ can be implemented efficiently in this diagonalized basis using the methods discussed in \appref{\ref{sec:appendix_circuits}}.

\subsubsection{Measurement and Readout of Circuits}
\label{sec:background_measurement_readout}

\begin{figure}[H]
    \centering

    \includegraphics[width=\linewidth]{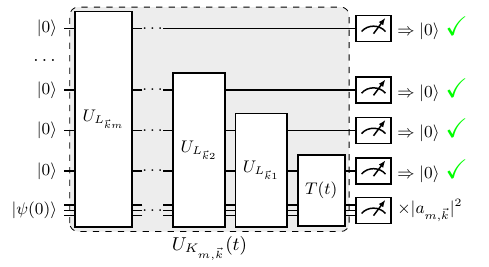}
    
    \caption{Diagram of the quantum circuit $U_{K_{m,\vec{k}}}(t)$ corresponding to a single Kraus operator in the time-perturbative Kraus series Eq. \eqref{eqn:tp_kraus_series}. This circuit applies a set of unitary dilations of the Lindblad operators ($U_{L_{\vec{k}}}$) indexed by the vector $\vec{k}$. Then, the effective Hamiltonian evolution $T(t)$ is applied. When all of the ancillas are measured to read $\ket{0}$, the resulting measurement probabilities on the system qubits are re-scaled by the factor $|a_{m, \vec{k}}|^2$ and added on a classical computer.}
    \label{fig:kraus_circuit}
\end{figure}

For each operator Eq. \eqref{eqn:tp_kraus_operator} in the Kraus series expansion, the unitaries $U_{L_n}$ (implementing the Lindblad operators) and $T(t)$ (implementing the effective Hamiltonian evolution) can be combined to produce a circuit $U_{K_{m,\vec{k}}}(t)$ corresponding to each Kraus operator, as shown in \figref{\ref{fig:kraus_circuit}}. When each circuit is run on a quantum device, both the system qubits and all ancillas are measured. When the ancillas all read $\ket{0}$, the the resulting system qubits are measured in a desired basis and the estimated outcome probabilities for each circuit are added on a classical computer with scalar weights $|a_{m,\vec{k}}|^2$.

Measurements in multiple bases can be combined to estimate the expectation value of the observable for the state $\rho(t)$. The state density matrix $\rho(t)$ itself can also be estimated by performing quantum state tomography in a complete basis \cite{cai_optimal_2016}. On NISQ devices, the Pauli basis is often chosen because transformations into this basis require only single-qubit Pauli gates. Although the Kraus operator method requires many circuits to be evaluated per measurement basis (in the worst case, $M!$ circuits), we note that the depth of each order $m$ circuit is at most $O(m)$. Furthermore, the Kraus series terms that contribute the most probability mass to the total system trajectory tend to correspond to the shortest circuits. As a result, this approach naturally mitigates the detrimental effect of noise on NISQ devices by distributing the computational cost of the simulation over many short circuits.

\subsection{Special Cases}
\label{sec:background_special_cases}

Although the time-perturbative Kraus series method is limited to systems that satisfy conditions (i)-(iv) and can require many circuits in the most general case, we note that there are some important classes of systems where our method not only applies, but can be made very efficient. In some cases, the time-perturbative Kraus series can even be reduced to a finite number of terms, allowing for an exact simulation distributed over only a finite number of circuits. In the following subsections, we provide further analysis of these noteworthy special cases:
\subsubsection{Systems with Unitary Symmetry}
\label{sec:background_unitary_cases}

First, we examine the special case of systems that exhibit unitary symmetries. These systems have a Hamiltonian $H$ and unitary operators $U_1, U_2, ..., U_N$ that satisfy $U_n H U_{n}^{\dagger} = H$. 
The unitaries $U_n$ can be used as Lindblad operators $L_n = U_n$ with appropriate choice of damping parameters $\gamma_n$ to phenomenologically model decoherence that drives the system toward a steady state that is invariant under these symmetries. For these systems it can be shown that conditions (i)-(iv) are satisfied with constants $\nu = \lambda = 0$, resulting in $f(t) = t$. The effective Hamiltonian evolution operator $T(t)$ takes the form
\begin{equation}
    T(t) = e^{-\Gamma t/2} U_H(t)
\end{equation}
where $\Gamma = \sum_{n=1}^N \gamma_n$ and $U_H(t) = \exp(-it H/\hbar)$ is the time evolution operator under the regular system Hamiltonian $H$. The scalar prefactor $e^{-\Gamma t/2}$ can be computed classically, which means that one only needs to apply the sequence of unitaries $U_H(t)U_{\vec{k}_1}U_{\vec{k}_2}...U_{\vec{k}_N}\ket{\psi}$ to simulate each Kraus term on a quantum computer.

\subsubsection{Bosonic Systems}
\label{sec:background_bosonic_cases}

Another special case of systems we consider here are bosonic systems. The Hamiltonian of a free bosonic system with a single mode can be modeled as a single-particle harmonic oscillator with Hamiltonian
\begin{equation}
    H = \hbar\omega \hat{a}^{\dagger}\hat{a}
    \label{eqn:harmonic_oscillator}
\end{equation}
where $\hat{a}$ is the mode annihilation operator and $\omega$ is the mode frequency. To model particle loss in the mode, we apply the Lindblad operator $L_1 = \hat{a}$. It can be shown that this system satisfies conditions (i)-(iv) with constants $\nu = \hbar\omega$ and $\lambda = -\gamma_1$, which means that $\alpha = \gamma_1 \ge 0$. Thus, the time-dependence of the the derived time-perturbative Kraus series is of the form $f(t) = (1-e^{-\alpha t})/\alpha$, meaning the Kraus series converges rapidly as claimed earlier. We can generalize the single mode case to a quadratic bosonic Hamiltonian with $n$ coupled modes, which can be written as
\begin{equation}
    H = \sum_{i,j} M_{ij} \hat{a}_i^{\dagger}\hat{a}_j
    \label{eqn:bosonic_hamiltonian}
\end{equation}
where $M$ is an $n \times n$ Hermitian matrix and $\hat{a}_i$ is the annihilation operator for mode $i$. In a manner similar to the Hamiltonian, Lindblad operators can be introduced as a matrix quadratic form
\begin{equation}
    L_n = \sum_{i,j} (V_n)_{ij} \hat{a}_i^{\dagger}\hat{a}_j
    \label{eqn:bosonic_lindblad}
\end{equation}
where $V_n$ is an arbitrary matrix. As a direct consequence of the Jordan-Schwinger map, the commutator algebra of $H$ and the Lindblad operators $L_n$ is isomorphic to that of the matrices $M$ and $V_n$. As a result, one only needs to consider the commutation relations of these matrices when verifying that conditions (i)-(iv) above apply. An important case to consider is when the matrices $V_n$ are diagonal or correspond to unitary symmetries of $M$ (i.e. $V_n M V_n^{\dagger} = M$). These kinds of systems are relevant in the theory of quantum angular momentum \cite{schwinger_angular_1952}, coupled optical cavities \cite{fabre_modes_2020}, and phonons in solids \cite{bolmatov_unified_2015}.

We would like to emphasize that the damped harmonic oscillator systems that can be simulated using the Trotterless method can be generalized to model the decay of quadratic bosonic systems under a Bogoliubov transformation \cite{xiao_theory_2009}. This would allow for a Hamiltonian of the form 
\begin{equation} H=\sum_{i,j}M_{ij}\hat{a}_i^{\dagger}\hat{a}_j+\frac{1}{2}\left(\Delta_{jk}\hat{a}_j^{\dagger}\hat{a}_k^{\dagger}+\Delta_{jk}^*\hat{a}_j\hat{a}_k\right) \label{eq:quad}\end{equation} 
where $M$ is a Hermitian matrix, $\Delta$ is a symmetric matrix, and the Lindblad operators are in terms of the transformed bosonic annihilation operator $L_i=\hat{a}_i$. 
This Hamiltonian in Eq. \eqref{eq:quad} can realize the four wave-mixing/spontaneous parametric down conversion Hamiltonian \cite{Vendromin_2022} in a lossy cavity. This is particularly relevant in quantum imaging when trying to obtain strong squeezing in leaky cavities \cite{seifoory_2017}. 
There has also been interest in using Lindbladians with dissipative couplings to engineer non-abelian states in photonic lattices \cite{Parto_2023}. 
The case where $\Delta_{jk}=0$ can be related to the dissipative evolution of various spin and fermionic Hamiltonians in the single-excitation subspace \cite{Sturges_2021}, as long as the Lindblad operators are particle/excitation number conserving. 

\subsubsection{Lindblad Operators with Finite Group and Semigroup Structure}
For systems with Lindblad operators that form a finite group or semigroup, this infinite series can be simplified down to a finite number of Kraus operators, yielding an exact solution. Of these types of systems, the most important case to consider is when the action of the Lindblad operators $L_n$ on some initial density matrix $\rho$ form finite abelian groups or finite abelian nilpotent semigroups, meaning they satisfy
\begin{equation}
    L_{n'}L_n \rho L_n^{\dagger}L_{n'}^{\dagger} = L_nL_{n'} \rho L_{n'}^{\dagger}L_n^{\dagger} 
    \label{eqn:lindblad_commutativity}
\end{equation}
(i.e. the action of the Lindblad operators on $\rho$ is commutative) and
\begin{equation}
    (L_n)^{\ell_n}\rho (L_n^{\dagger})^{\ell_n} = \theta_n \rho
    \label{eqn:lindblad_power_law}
\end{equation}
for real constants $\theta_n \geq 0$ and integers $\ell_n > 0$. If $\theta_n = 0$, the group generated by the action of $L_n$ on $\rho$ is a finite nilpotent semigroup; otherwise it forms a finite abelian group (up to multiplication by some constant). In either case, it can be shown that the resulting time-perturbative Kraus series can be reduced to the finite sum of Kraus operators
\begin{equation}
    \rho(t) = \sum_{m_1=0}^{\ell_1-1} \sum_{m_2 = 0}^{\ell_2-1} ... \sum_{m_N = 0}^{\ell_N -1} K_{\vec{m}}(t)\rho(0)K_{\vec{m}}(t)^{\dagger}
    \label{eqn:reduced_tp_kraus_series}
\end{equation}
where
\begin{equation}
    K_{\vec{m}}(t) = T(t)\prod_{n=1}^N \sqrt{F_{\ell_n,{m}_n}^{\theta_n}(\gamma_n t)}(L_n)^{m_n}
\end{equation}
and $F_{\ell,m}^{\theta}$ is the generalized hyperbolic function
\begin{equation}
    F_{\ell,m}^{\theta}(x) = \frac{1}{\ell}\theta^{-m/\ell} \sum_{k=0}^{\ell-1} \omega_{\ell}^{-mk} \exp(\omega_{\ell}^{k}\theta^{\frac{1}{\ell}}x)
\end{equation}
with $\omega_\ell = e^{i2\pi/\ell}$. In the nilpotent case ($\theta = 0$), we define $F_{\ell, m}^0(x) = f(x)^m/(m!)$.

\begin{figure}[h!]
    \centering

    \begin{tabular}{p{0.95\linewidth}}
        \lsubfigimg[width=0.75\linewidth]{(a)}{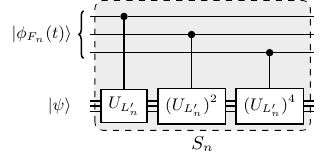} \\
        \lsubfigimg[width=\linewidth]{(b)}{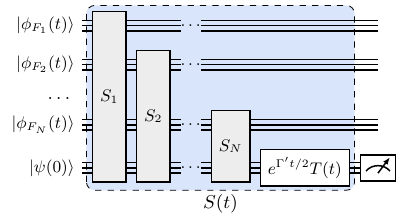}
    \end{tabular}
    \caption{Quantum circuit diagrams for the circuits $S_n$ (a) and $S(t)$ (b), which can be used to efficiently simulate systems where the action of the Lindblad operators that form a finite abelian group. Simulating these systems requires the hyperbolic distributions $\phi_{F_n}(t)$ to be prepared on sets of ancilla qubits.}
    \label{fig:factored_group_circuit}
\end{figure}

Although the finite sum of Kraus operators in Eq. \eqref{eqn:reduced_tp_kraus_series} contains many individual terms, we observe that this series can be factored as the product of polynomial functions of the Lindblad operators. Specifically, in the finite abelian case with $\theta_n \neq 0$, we can re-scale the Lindblad operators via the transformation in Eq. \eqref{eqn:lindblad_rescaling} with $b_n = \theta_n^{1/(2\ell_n)}$. This allows us to write the time evolution of the system as $\rho(t) =  S(t)\rho S(t)^{\dagger}$ where $S(t)$ is given by
\begin{equation}
    \begin{aligned}
    S(t) &= e^{\Gamma' t/2}T(t) \\
    &\qquad \times \prod_{n=1}^N\left( \sum_{m = 0}^{\ell_n-1}  \sqrt{e^{-\gamma_n' t} F_{\ell_n, m}^{1}(\gamma_n' t)} (L_n')^{m} \right)
    \end{aligned}
    \label{eqn:evolution_S}
\end{equation}
where $L_n' = L_n/\theta^{1/(2\ell_n)}$, $\gamma_n'= \theta^{1/\ell_n}\gamma_n$, and $\Gamma' = \sum_{n=1}^N \gamma_n'$. If one can efficiently apply a controlled form of the gates $U_{L_n'}$ that implement each re-scaled Lindblad operator $L_n'$, then each factor in Eq. \eqref{eqn:evolution_S} can be applied using circuits of the form shown in \figref{\ref{fig:factored_group_circuit}}, provided that the distribution
\begin{equation}
    \ket{\phi_{F_n}(t)} = \sum_{m=0}^{\ell_n-1} \sqrt{e^{-\gamma_n' t} F_{\ell_n,m}^{1}(\gamma_n' t)}\ket{m}
\end{equation}
can be efficiently prepared for each time evolution period $t > 0$ on a set of $\log_2(\ell_n)$ ancilla qubits. It can be shown that a circuit preparing the distribution $\ket{\phi_{F_n}(t)}$ can be compiled and executed efficiently, requiring at most $O(\ell_n \log(\ell_n))$ quantum gates and classical preprocessing steps \cite{araujo_divide-and-conquer_2021}. 

\section{Results}
\label{sec:results}

To demonstrate the effectiveness of our proposed framework on NISQ hardware, we present results of simulations of basic open quantum systems on noisy trapped ion quantum computers, specifically on the IonQ Harmony \cite{noauthor_ionq_2024} and Quantinuum H1-1 \cite{noauthor_quantinuum_2024} devices. These devices have slightly higher noise in comparison to state-of the-art devices, but the noise syndromes of these devices are well-understood \cite{wright_benchmarking_2019}, making them a good platform for demonstrating NISQ-friendly algorithms and error mitigation strategies in high-throughput simulation tasks. Quantum circuits were designed and compiled via the qiskit software framework \cite{javadi-abhari_quantum_2024} for all IonQ results, and via the pytket framework \cite{Sivarajah_2021} for all Quantinuum device results.

To start, we present simulation results for the Pauli channel and damped Schwinger oscillator models, and discuss how these models satisfy the criterion of the special cases discussed in \secref{\ref{sec:background_special_cases}}. We then present results obtained from simulating the trajectory of the system's density matrix on real quantum hardware. Next, we focus on simulating a damped quantum harmonic oscillator (i.e. a single bosonic mode occupied by many particles), and demonstrate how error mitigation techniques can be applied to suppress the effect of noise on NISQ devices.

\subsection{Continuous-Time Pauli Channel}
\label{sec:ct_pauli_channel}

The continuous-time Pauli channel is a type of open quantum system used to model the noise induced in a quantum computing device as a function of time \cite{david_digital_2024}, or to model decoherence in degenerate spin-1/2 states coupled to a spin bath \cite{bhattacharya_exact_2017}. In most cases, it is common to assume a trivial Hamiltonian $H = 0$, with unitary Lindblad operators $L_n$ with respective error rates $\gamma_n$ that generate an error syndrome on $N_q$-qubit states in the device. These Lindblad operators assume the form of a Pauli string
\begin{equation}
    L_n = \sigma_{n1} \otimes \sigma_{n2} \otimes ... \otimes \sigma_{n N_q},
\end{equation}
where each $\sigma_{ij} \in \{ I, X, Y, Z \}$ is a single-qubit Pauli operator and $N_q$ is the number of qubits. Since $H = 0$ and the operators $L_n$ are unitary, we have $[H,L_n] = 0$, which means that the system satisfies conditions (i)-(iv) as discussed in \secref{\ref{sec:background_unitary_cases}}. In addition, it can be shown that Pauli string Lindblad operators form a finite abelian group that satisfies Eq. \eqref{eqn:lindblad_commutativity} and Eq. \eqref{eqn:lindblad_power_law} with $\ell_n = 2$ for each $L_n$. This means that the system admits a finite Kraus representation of the form \eqref{eqn:reduced_tp_kraus_series} with Kraus operators
\begin{equation}
    K_{\vec{m}}(t) = e^{-\Gamma t/2}\prod_{n=1}^N \sqrt{F_{2,{m}_n}^{1}(\gamma_n t)} (L_n)^{m_n}
    \label{eqn:pc_kraus_operators}
\end{equation}
which can be combined in superposition to obtain a factored time evolution operator
\begin{equation}
    S(t) = \frac{1}{\sqrt{2^N}}\prod_{n=1}^N\left( \sqrt{1+e^{-\gamma_n t}} I + \sqrt{1-e^{-\gamma_n t}} L_n \right).
\end{equation}
This operator can be efficiently simulated using a quantum circuit of the form shown in \figref{\ref{fig:factored_group_circuit}}. Since $\ell_n = 2$, only a single ancilla qubit is needed to prepare the distribution $\ket{\phi_{F_n}(t)}$ corresponding to each Lindblad operator and each evolution period $t$. This circuit requires one two-qubit gate for each Pauli component $\sigma_{ij} \neq I$ in the set of Lindblad operators. Alternatively, one can also simulate a Pauli channel using the non-factored Kraus operators in Eq. \eqref{eqn:pc_kraus_operators}, making the observation that the time-dependent terms $e^{-\Gamma t/2}\prod_{n} \sqrt{F_{2,m_n}^1(\gamma_n t)}$ can be applied on a classical computer after measuring the initial state under the action of the powers $L_n^m$ of each Kraus operator. This requires performing partial (or in the worst case, a full) tomography on the initial state; however, the result of these measurements can be re-combined classically to find the continuous system trajectory for all values of $t > 0$.

\begin{figure}
    \centering
     \begin{tabular}{p{\linewidth}}
          \subfigimg[width=\linewidth]{(a)}{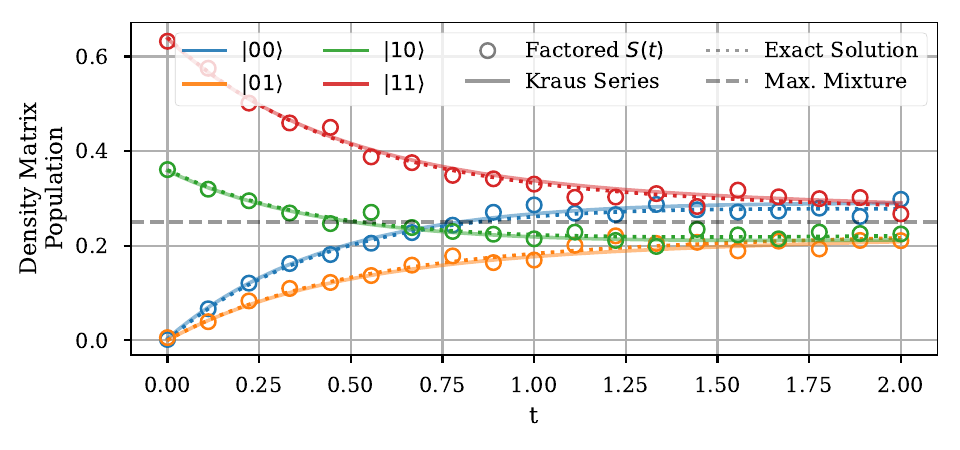} \\ 
          \subfigimg[width=\linewidth]{(b)}{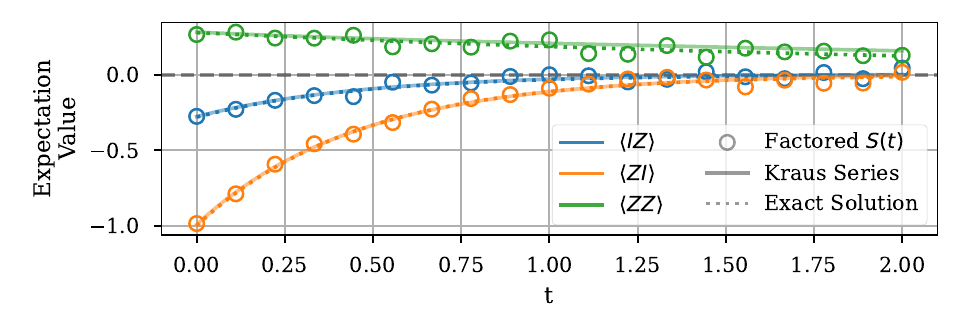} 
     \end{tabular}
    \caption{Trajectory of a two-qubit Pauli channel obtained from the noisy Quantinuum H1-1 device. Results are shown for (a) the populations along the diagonal of the density matrix and (b) the expectation values of Pauli strings in the $Z$-basis. For each plot, trajectories using the standard Kraus operator method (solid lines) and the factored $S(t)$ operator (circles) are shown.}
    \label{fig:pauli_channel_results}
\end{figure}

As a motivating demonstration, we simulate a two-qubit Pauli channel under strongly correlated $XX$ and $ZZ$ interactions. This specific Pauli channel model is relevant in modeling qubit crosstalk in trapped ion quantum processors. For this system we use the Lindblad operators
\begin{equation}
\begin{aligned}
        L_1 = I \otimes X, \quad & L_2 = X \otimes I, \\
        L_3 = Z \otimes Z, \quad &  L_4 = X \otimes X
\end{aligned}
\label{eqn:pc_lindblad_operators}
\end{equation}
with single-qubit error rates $\gamma_1,\gamma_2 = 0.1$ and two-qubit error rates $\gamma_3, \gamma_4 = 1.0$. We simulated the trajectory of the initial unentangled pure state
\begin{equation}
    \ket{\psi(0)} = -3/5\ket{01} - 4/5\ket{11}
\end{equation}
on the noisy Quantinuum H1-1 device, using both the $S(t)$ and the full Kraus series method. In \figref{\ref{fig:pauli_channel_results}}, we plot the trajectory of the diagonal populations of the density matrix and the $Z$-basis Pauli operators estimated with $2048$ shots per circuit. These results are compared with an exact numerical solution of the Lindblad equation. 

Due to the strongly correlated nature of the simulated Pauli errors, we observe in \figref{\ref{fig:pauli_channel_results}(a)} that the system is driven from an unentangled pure state at $t = 0$ to a mixed, yet weakly correlated state at $t = 2.0$. While the single-qubit polarizations in the $Z$-basis decay to zero in \figref{\ref{fig:pauli_channel_results}(b)}, the $ZZ$-correlation between the qubits decays at a much slower rate. 
In \figref{\ref{fig:pauli_channel_results}}, we also note the strong agreement between the Kraus series solution and the exact solution to the Lindblad equation. We also observe strong agreement between the results of the factored $S(t)$ method and the exact solution, but with increased noise due to the larger number of two-qubit gates used in the $S_n$ circuits depicted in \figref{\ref{fig:factored_group_circuit}}.

\subsection{Damped Schwinger Oscillator Model}
\label{sec:schwinger_oscillator}

The Schwinger Oscillator model is a 2D quantum harmonic oscillator system that can be equivalently viewed as a degenerate bosonic mode with two degrees of freedom. This model was first proposed by Julian Schwinger in 1952 to explain the theory of quantum angular momentum with two uncoupled oscillators \cite{schwinger_angular_1952}. A degenerate 2D oscillator has the diagonalized Hamiltonian
\begin{equation}
    H = \hbar\omega(1 + \hat{a}_1^{\dagger}\hat{a}_1 + \hat{a}_2^{\dagger}\hat{a}_2).
    \label{eqn:schwinger_hamiltonian}
\end{equation}
Up to addition of scalar terms, this Hamiltonian can be written as a quadratic product of bosonic creation and annihilation operators of the form in Eq. \eqref{eqn:bosonic_hamiltonian} with the matrix $M = \hbar\omega I$, where $I$ is the $2\times2$ identity. Following the analysis in \secref{\ref{sec:background_bosonic_cases}}, we can efficiently simulate quadratic Lindblad operators of the form in Eq. \eqref{eqn:bosonic_lindblad}, provided that the associated matrices $M$ and $V_n$ satisfy conditions (i)-(iv) in place of $H$ and $L_n$ respectively. 
For the $2$-mode case, a natural choice of $V_n$ is the set of Pauli operators ($X, Y, Z$), which generate $SU(2)$. In their bosonic quadratic form, these are proportional to the quantum angular momentum operators, as first proposed by Schwinger:
\begin{align}
    \hat{J}_x &= (\hat{a}_1^{\dagger}\hat{a}_2 + \hat{a}_2^{\dagger}\hat{a}_1)/2 
        \label{eqn:schwinger_Jx} \\
    \hat{J}_y &= (\hat{a}_1^{\dagger}\hat{a}_2 - \hat{a}_2^{\dagger}\hat{a}_1)/2i
        \label{eqn:schwinger_Jy} \\
    \hat{J}_z &= (\hat{a}_1^{\dagger}\hat{a}_1 - \hat{a}_2^{\dagger}\hat{a}_2)/2
        \label{eqn:schwinger_Jz}
\end{align}
Since $M = \hbar\omega I$ and the Pauli operators $V_n \in \{X, Y, Z\}$ meet the conditions for satisfying (i)-(iv), we can simulate the evolution of the system under a subset of the Lindblad operators Eqs. \eqref{eqn:schwinger_Jx}-\eqref{eqn:schwinger_Jz} using a time-perturbative Kraus series of the form in Eq. \eqref{eqn:tp_kraus_series} with $f(t) = t$.

In accordance with convention, the $\hat{J}_z$ operator is chosen to be diagonal, though we recall that an arbitrary angular momentum operator $\hat{J}_{\hat{n}}$ about an axis $\hat{n}$ is similar to $\hat{J}_z$ under the unitary transformation $\exp(-i\pi J_{\hat{v}})$, where $\hat{v} = (\hat{n} + \hat{z})/\norm{\hat{n} + \hat{z}}$. As a result, one can always choose at least one Lindblad operator to have the diagonalized representation $\hat{J}_{\hat{z}}$, which can be applied more efficiently on a quantum computer.
When a particular angular momentum operator (e.g. $\hat{J}_z$) is applied as a Lindblad operator, it produces a depolarization in the angular momentum of the oscillator about the given axis while conserving the total energy of the system. This can model, for example, the decoherence experienced by a charged particle occupying degenerate quantized Landau levels under a strong uniform magnetic field. (In this particular example, one recovers the 2D oscillator representation when the magnetic field is treated in the symmetric gauge.) The quantization of these degenerate Landau levels is relevant in explaining the Shubnikov-de Haas effect \cite{laikhtman_quasiclassical_1994} and the integer quantum Hall effect \cite{hatsugai_chern_1993}.

To demonstrate the application of our methods to this system, we simulate a 2D oscillator with a single occupied doubly-degenerate Landau orbit level with energy $E_L = 2\hbar\omega$. Simulating up to the first degenerate level requires two qubits, where the degenerate subspace is spanned by the logical states $\ket{01}$ and $\ket{10}$. We apply angular damping $L_1 = J_z$ to an initially pure combination of the ground and excited states
\begin{equation}
    \ket{\psi(0)} = (\ket{00} + i\ket{01} + \ket{10})/\sqrt{3}.
\end{equation}
and estimate the trajectory of the full density matrix by performing complete Pauli-basis tomography on the system qubits. Quantum simulations were carried out on the Quantinuum H1-1 device to reconstruct the full system density matrix at five different points in time with 1024 shots per circuit.

\begin{figure}
    \centering
     \begin{tabular}{p{\linewidth}}
        \subfigimg[width=0.96\linewidth]{(a)}{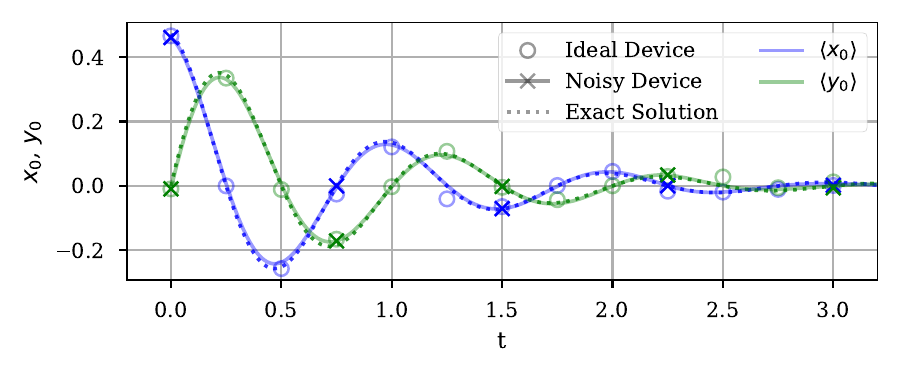}\\
        \subfigimg[width=0.96\linewidth]{(b)}{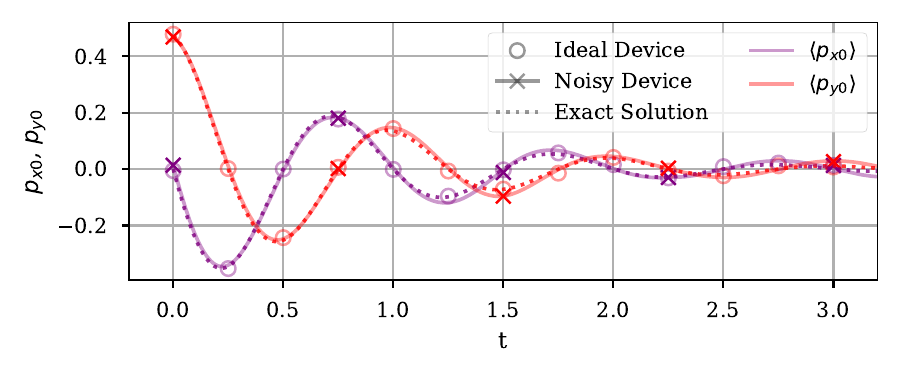} \\
        \subfigimg[width=\linewidth]{(c)}{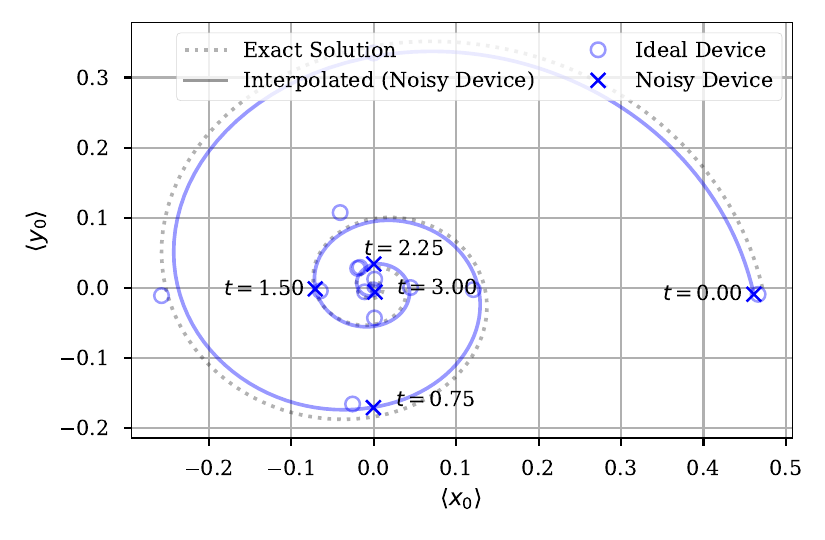}
     \end{tabular}
    \caption{Trajectory of the expectation values of position and momentum for oscillator 1 ($x_0,p_{x0}$) and oscillator 2 ($y_0, p_{y0}$) computed on the noisy Quantinuum H1-1 device. Trajectories are shown for (a) the scaled position operators (b) the scaled momentum operators, and (c) for $(x_0,y_0)$ joint spatial coordinates. The noisy results are compared to both an ideal quantum computation (containing only sampling error) and the exact solution to the Lindblad equation. The solid lines in plots (a)-(c) show an interpolation of the noisy data corresponding to the trajectory of a damped classical harmonic oscillator model with frequency $\omega$.}
    \label{fig:schwinger_trajectory}
\end{figure}

In \figref{\ref{fig:schwinger_trajectory}} we plot the expectation value of the trajectory of the system in mass-independent coordinates using the operators $x_0 = (\hat{a}_1^\dagger + \hat{a}_1)/\sqrt{2}, y_0 = (\hat{a}_2^\dagger + \hat{a}_2)/\sqrt{2}$ and their momentum analogues $p_{x0} = i(\hat{a}_1^{\dagger} - \hat{a}_1)/\sqrt{2}, p_{y0} = i(\hat{a}_2^{\dagger} - \hat{a}_2)/\sqrt{2}$, where $\hat{a}_1, \hat{a}_2$ are restricted to act on only the lowest modes $\ket{0}$ and $\ket{1}$. For both the position and momentum coordinates, we observe strong agreement between the exact solution and the estimated values (for both the noisy H1-1 device and a simulated ideal noiseless device). 

\begin{figure}
    \centering
     \begin{tabular}{p{\linewidth}}
          \subfigimg[width=\linewidth]{(a)}{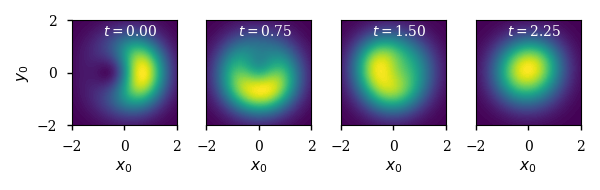} \\[-0.3cm]
          \subfigimg[width=\linewidth]{(b)}{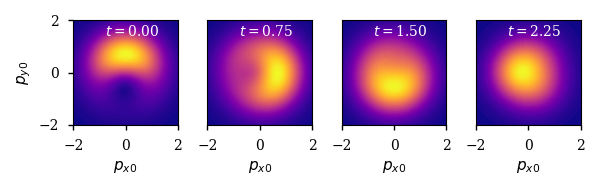}
     \end{tabular}
    \caption{Snapshots of the trajectory of the joint position (a) and momentum (b) density of the Schwinger oscillator system. These densities were estimated directly from the density matrices obtained from the noisy Quantinuum H1-1 device.}
    \label{fig:schwinger_snapshots}
\end{figure}

As $t \rightarrow \infty$, the position and momentum coordinates asymptotically approach zero, which is consistent with the depolarization of angular momentum in the $z$-direction. The depolarization can be visualized by reconstructing the full position and momentum distributions from the density matrix estimated on the noisy device. We expand the density matrix $\rho_{in,i'n'}$ in the position and momentum basis as a sum of products of Hermite-Gauss polynomials, given by
\begin{equation}
    \begin{aligned}
    \rho(\vec{x_0}) &= \sum_{n,n'}\prod_{i,i'} \Big{(} \rho_{in,i'n'} \\
    &\qquad \times\frac{\exp({-(\vec{x_0}_i)^2/2})}{\sqrt{2^{n+n'}\pi (n!)(n'!)}} H_n(\vec{x_0}_i)H_{n'}(\vec{x_0}_{i'}) \Big{)}
    \end{aligned}
    \label{eqn:position_density}
\end{equation}
where $i,i'$ are the coordinate (i.e. mode) indices and $n,n'$ are the coordinate eigenstate indices. The analogous momentum density can be obtained through a Fourier transform of Eq. \eqref{eqn:position_density}. In \figref{\ref{fig:schwinger_snapshots}} we plot the position and momentum density estimated on the Quantinuum H1-1 device for the first four points in time.

\subsection{Damped Harmonic Oscillator with Error Mitigation}
\label{sec:qho_error_mitigation}

So far, we have simulated relatively small systems on NISQ hardware, where the effect of noise is on the same order of magnitude as the sampling error. However, when simulating much larger systems on NISQ devices, the effect of noise is more problematic and must be combated through techniques such as hardware-native gate optimization and error mitigation \cite{sun_mitigating_2021, cai_quantum_2023}. In this section, we focus on error-mitigation techniques, as circuit optimization is more hardware specific. In particular, we propose hardware-agnostic error mitigation techniques that can be applied to improve the accuracy of the density matrix trajectories of bosonic systems simulated using the Kraus series method. We apply these techniques to a damped harmonic oscillator system with Hamiltonian Eq. \eqref{eqn:harmonic_oscillator} and Lindblad operator $L_1 = \hat{a}$ with damping parameter $\gamma_1$. We restrict the representation of $\hat{a}$ to incorporate occupied eigenstates up to $n=3$ in a dense two-qubit binary encoding. This allows for a Kraus representation of the form in Eq. \eqref{eqn:tp_kraus_series} with $f(t) = (1-e^{-\gamma_1 t}))/\gamma_1$.

As with the Schwinger oscillator system, we estimate the trajectory of the two-qubit oscillator density matrix by performing full tomography in the Pauli basis. While this process of reconstructing the density matrix can be expensive (requiring measurement in nine different bases for each Kraus operator circuit), it provides a significant amount of data that can be used for error mitigation both at and above the hardware level. However, when simulating open quantum systems, the system density matrix is not always pure, which is guaranteed to be the case when considering the unitary quantum evolution of a closed system on a noiseless device. Furthermore, the decoherence of the system qubits due to entanglement with ancilla qubits combines with the device noise acting on the system qubits to produce a state that is often more decoherent and has a larger entropy than desired. This problem of distinguishing the simulated noise from the device noise poses a unique challenge that remains to be satisfactorily addressed in the literature.

One simple method for mitigating device noise is to model the noise as some CPTP channel $\mathcal{E}(\rho)$ and attempt to partially mitigate device noise by inverting the channel. The inverted channel is applied to the noisy density matrix $\rho$ to obtain a mitigated density matrix $\rho_{\text{mit}}$. As we have seen in \secref{\ref{sec:ct_pauli_channel}}, one such model is given by a Pauli channel. The evolution of a density matrix $\rho$ for a constant time period under a Pauli channel with Lindblad operators $L_n$ of the form in Eq. \eqref{eqn:pc_lindblad_operators} can be always written in the simplified form
\begin{equation}
    \mathcal{E}_{\text{PC}}(\rho) = \varepsilon_0 \rho + \sum_{i} \varepsilon_i P_i \rho P_i^{\dagger},
    \label{eqn:pauli_channel_model}
\end{equation}
where $P_i \in \{I, X, Y, Z\}^{\otimes N} \setminus \{ I^{\otimes N}\}$ are non-trivial Pauli strings and $\varepsilon_i \in [0,1]$ are the associated error probabilities that sum to unity. An important special case of the general Pauli channel is the quantum depolarizing channel (QDC), where the error probabilities $\varepsilon_i$ for $i > 0$ are identical across the set of all $P_i$. Letting $\lambda = 1-\varepsilon_0$, it can be shown that the Pauli channel simplifies to
\begin{align}
    \mathcal{E}_{\text{QDC}}(\rho) &= (1-\lambda)\rho + \lambda\left(\sum_{P_i \in \mathcal{P}} P_i \rho P_i^{\dagger}\right) \\
    &= (1-\lambda)\rho + \frac{\lambda}{2^N}I^{\otimes N}.
    \label{eqn:depolarizing_channel_model}
\end{align}
One important property of the QDC is that the noise induced by this channel commutes with any simulated CPTP noise channel $\mathcal{E}_{\text{sim}}$, that is
\begin{equation}
    \mathcal{E}_{\text{sim}}(\mathcal{E}_{\text{QDC}}(\rho)) = \mathcal{E}_{\text{QDC}}(\mathcal{E}_{\text{sim}}(\rho))
    \label{eqn:depolarizing_channel_commutativity}
\end{equation}
for all density matrices $\rho$. This means that one can correct for device-induced depolarizing errors that occur before and during the simulation by applying an inverse depolarization channel to the measured density matrix after simulation is complete. Moreover, inverting $\mathcal{E}_{\text{QDC}}$ can be carried out quite efficiently, requiring only a re-scaling of the diagonal and off-diagonal components of $\rho$ to obtain the mitigated density matrix $\rho_{\text{mit}}$. Although the Pauli channel model in Eq. \eqref{eqn:pauli_channel_model} is a much more flexible noise model, it does not exhibit the same commutativity as in Eq. \eqref{eqn:depolarizing_channel_commutativity} except in the case when $\mathcal{E}_{\text{sim}}$ is also a Pauli noise model. For more information on the process of fitting QDCs and Pauli channels to measurements on quantum hardware, we refer the reader to \appref{\ref{sec:appendix_pauli_noise_models}}. In the remainder of this section, we discuss the results of two experiments that apply the error mitigation techniques discussed above to the damped harmonic oscillator system.

\subsubsection{Noisy Oscillating State}

First, we consider the simulation of the oscillating non-stationary state 
\begin{equation}
    \ket{\psi(0)} = \frac{1}{\sqrt{2}}(i\ket{2} + \ket{3}).
\end{equation}
which we simulated on the trapped ion IonQ Harmony device with damping rate $\gamma_1 = 1.0$ and $\hbar\omega = 1.0$ using the Kraus series method truncated to order $m = 3$. The full density matrices were estimated via tomography in the Pauli bases for 19 time steps uniformly spaced in the interval $t \in [0,3]$ with 1024 shots per circuit.

\begin{figure*}
    \centering
    \begin{tabular}{p{0.45\linewidth} p{0.45\linewidth}}
        \subfigimg[width=\linewidth]{(a)}{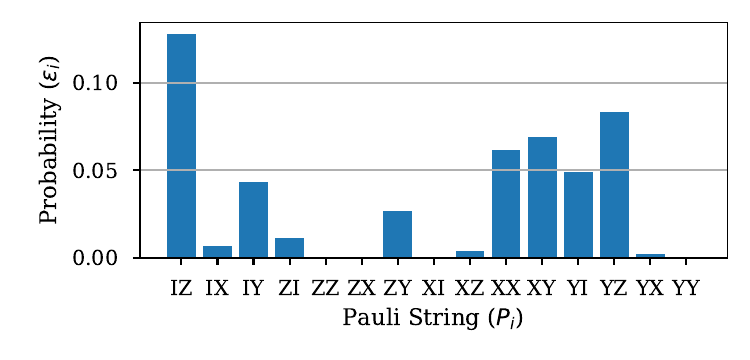} & \subfigimg[width=\linewidth]{(c)}{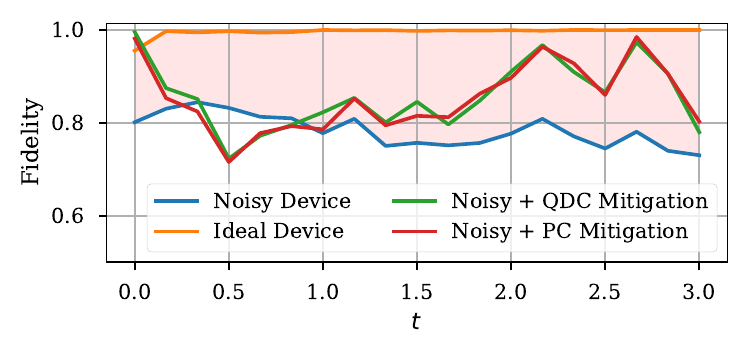} \\
        \subfigimg[width=\linewidth]{(b)}{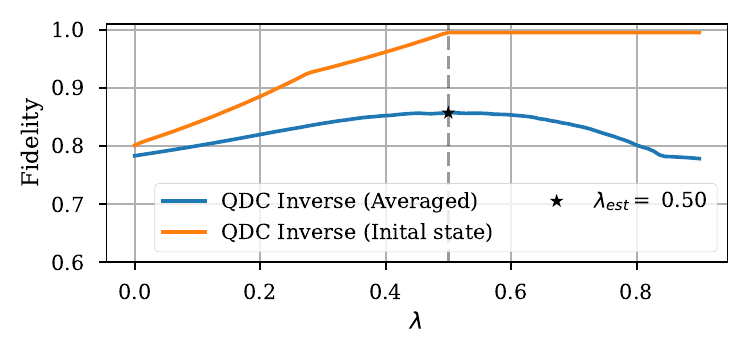} & 
        \subfigimg[width=\linewidth]{(d)}{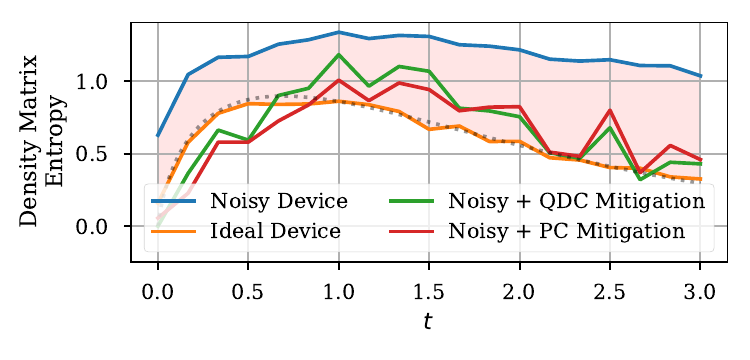}
    \end{tabular}
    \caption{(a): Pauli channel model fit to the device-induced noise on the IonQ Harmony device.
    (b): Effect of the quantum depolarizing channel (QDC) parameter $\lambda$ on the fidelity of the error-mitigated density matrix obtained from inverting a QDC fit to measurements obtained from the IonQ Harmony device. The average fidelity across the entire trajectory and only the prepared initial state are shown as a function of $\lambda$, suggesting an optimal value of $\lambda \approx 0.5$ for this device and system.\\
    (c,d): Fidelity (c) and von Neumann entropy (d) of the density matrix trajectory simulated on the IonQ Harmony device. For each time value $t$, results are shown for an ideal device (only sampling error), the noisy device, and the noisy device with error mitigation via inversion of a general Pauli channel and a QDC fit to the initial state only. }
    \label{fig:qho_ionq_qdc}
\end{figure*}

To mitigate the effect of noise decoherence on the Harmony device, we fit a Pauli channel model and a QDC model to the initial state and select the respective parameters $\varepsilon_i$ and $\lambda$ such that the fidelity of the density matrix reconstructed on the $t = 0$ density matrix is maximized. The results of the estimated Pauli channel and QDC parameters are shown in \figref{\ref{fig:qho_ionq_qdc}(a) and \ref{fig:qho_ionq_qdc}(b)} respectively. Since the QDC satisfies Eq. \eqref{eqn:depolarizing_channel_model}, an estimate of the parameter $\lambda$ that is sufficient to at least correct for decoherence during state preparation and measurement can be obtained by selecting the smallest value of $\lambda$ that maximizes the fidelity of the initial density matrix. As shown in \figref{\ref{fig:qho_ionq_qdc}(b)}, estimating $\lambda$ in this manner approximately maximizes the average fidelity of the mitigated density matrices across the entire trajectory. In \figref{\ref{fig:qho_ionq_qdc}(c) and \ref{fig:qho_ionq_qdc}(d)}, we plot the fidelity and von Neumann entropy of the density matrices estimated on the noisy device. In these figures, the gap between the noisy device results and a simulated ideal device are shaded in red, illustrating the effect of device noise decoherence. We observe in \figref{\ref{fig:qho_ionq_qdc}(d)} that both the Pauli channel and QDC error models decrease the density matrix entropy to closer resemble that of an ideal device. These methods also increase the average fidelity of the reconstructed density as seen in \figref{\ref{fig:qho_ionq_qdc}(d)}, especially in the regime where the entropy of the density matrix is low. On the other hand, in the high-entropy regime (where $t \approx 0.5$) both error mitigation techniques reduce the entropy below that of the exact solution, resulting in a lower density matrix fidelity than the raw unmitigated data. This is illustrative of a key trade-off that exists between the fidelity of the mitigated density matrix in the high-entropy regime and low-entropy regimes: as the degree of mitigation is increased (i.e. the noise channel parameters $\varepsilon_i$ and $\lambda$ are increased), the fidelity in the high-entropy regime increases at the cost of some fidelity of the lower-entropy states. We remark that the exact nature of this trade-off, however, is likely to be specific to both the system being simulated and the quantum hardware it is simulated on.

\begin{figure}
    \centering
    \begin{tabular}{c}
        \subfigimg[width=\linewidth]{(a)}{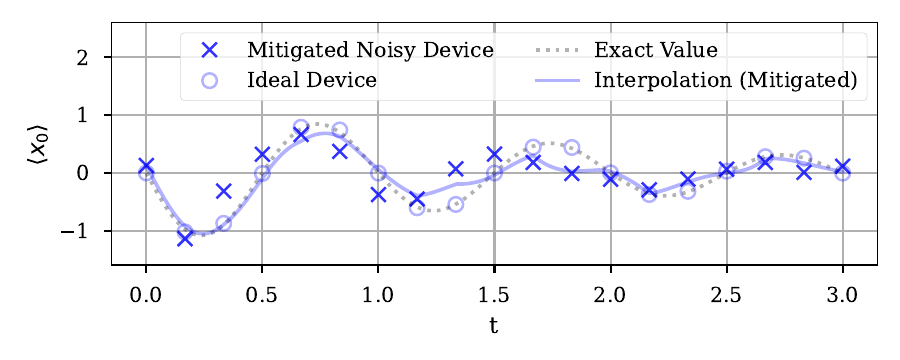} \\
        \subfigimg[width=\linewidth]{(b)}{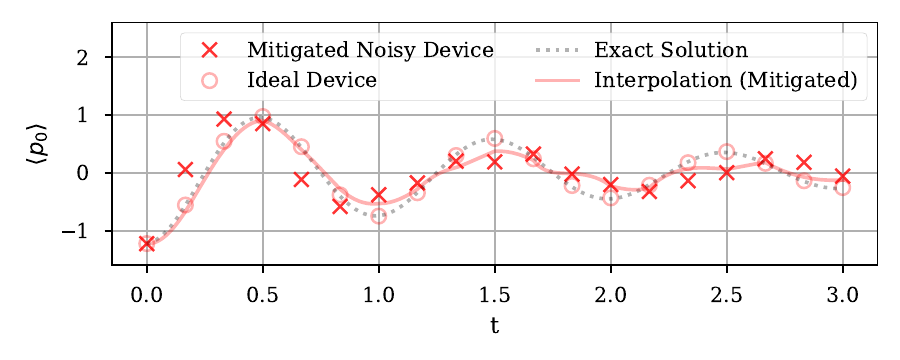}
    \end{tabular}
    \caption{Trajectory of error-mitigated expectation values of oscillator position (a) and momentum (b)  computed on the noisy IonQ Harmony device. The results for the mitigated density matrix, an ideal device, and the exact solution to the Lindblad equation are plotted for comparison. An interpolating trajectory of a classical harmonic oscillator with variable damping is also shown (solid line). This interpolation represents the least-squares error fit of the mitigated data.}
    \label{fig:qho_ionq_expvals}
\end{figure}

In \figref{\ref{fig:qho_ionq_expvals}} we plot the error-mitigated expectation value of position and momentum of the damped oscillator system, where we see reasonable agreement with the exact solution. This agreement can be made stronger by interpolating the mitigated position and momentum values using the model of a damped classical harmonic oscillator with a variable damping coefficient. This interpolation is represented by the solid lines in \figref{\ref{fig:qho_ionq_expvals}}.

In \figref{\ref{fig:qho_ionq_density_trajectories}}, we plot the trajectory of the position density and momentum density of the harmonic oscillator system interpolated as function of time. For each discrete time sample, the position and momentum densities were computed using Eq. \eqref{eqn:position_density} and interpolated between time steps with Gaussian smoothing. For comparison, the exact solution to the Lindblad equation is plotted in the last row of \figref{\ref{fig:qho_ionq_density_trajectories}}. Although the unmitigated results appear to be degraded with noise, we see that much of this noise is suppressed after the inverse QDC error mitigation is applied. Nonetheless, the effect of noise is still prominent in the high-entropy regime at $t \approx 1.0$. 

\begin{figure*}
    \centering
    \includegraphics[width=0.9\linewidth]{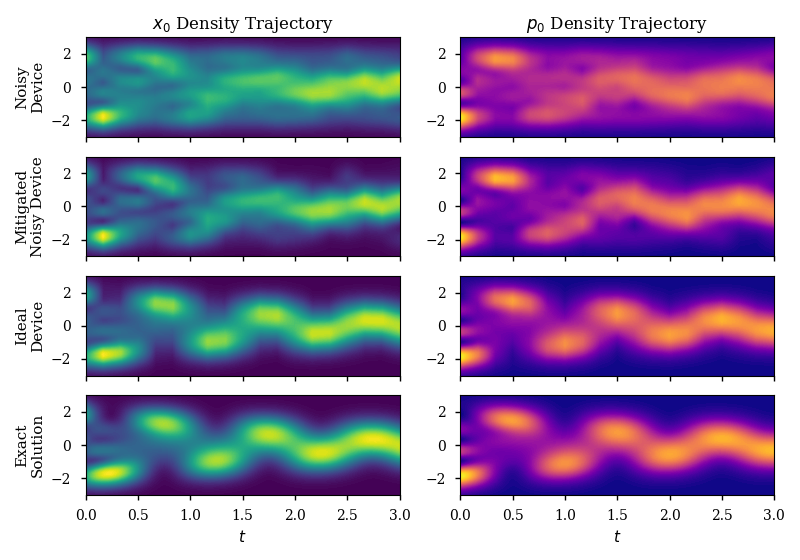}
    \caption{Trajectory of the position density (left column) and momentum density (right column) of the harmonic oscillator prepared in a non-stationary state. Densities are plotted for the noisy IonQ Harmony device results, the error-mitigated noisy results, an ideal quantum device trajectory, and finally the exact solution to the Lindblad equation. For the device results, the distributions between discrete time steps are estimated through Gaussian interpolation.}
    \label{fig:qho_ionq_density_trajectories}
\end{figure*}

As the oscillator evolves, the dynamics of the oscillator transition from behaving like a quantum-mechanical object, where measurements of position and momentum are governed by the uncertainty principle, to a classical statistical object, where measurement outcomes can be described by a joint probability function of position and momentum. This transition can be observed by computing the Wigner quasi-probability distribution of the harmonic oscillator as a function of time. The Wigner quasi-probability distribution of a one-dimensional system with density operator $\hat{\rho}$ is given by:
\begin{equation}
    W(x_0, p_0) = \frac{1}{\pi}\int \bra{x_0+s}\rho\ket{x_0-s} e^{i2 p_0 s}\ ds
    \label{eqn:wigner}
\end{equation}

The Wigner distribution plays a crucial role in the phase-space formulation of quantum mechanics and can be interpreted as a joint distribution over the position and momentum phase space of a system. Unlike a standard probability distribution, this quasi-probability distribution can sometimes take on negative values, which are attributed to inherently quantum-mechanical effects arising from the non-commutativity of $x_0$ and $p_0$. Furthermore, when the quasi-probability distribution is entirely positive, it is indicative of classical statistical behavior (e.g., an ensemble of Gaussian states) \cite{mandilara_extending_2009}.

\begin{figure*}
    \centering
    \includegraphics[width=0.86\linewidth]{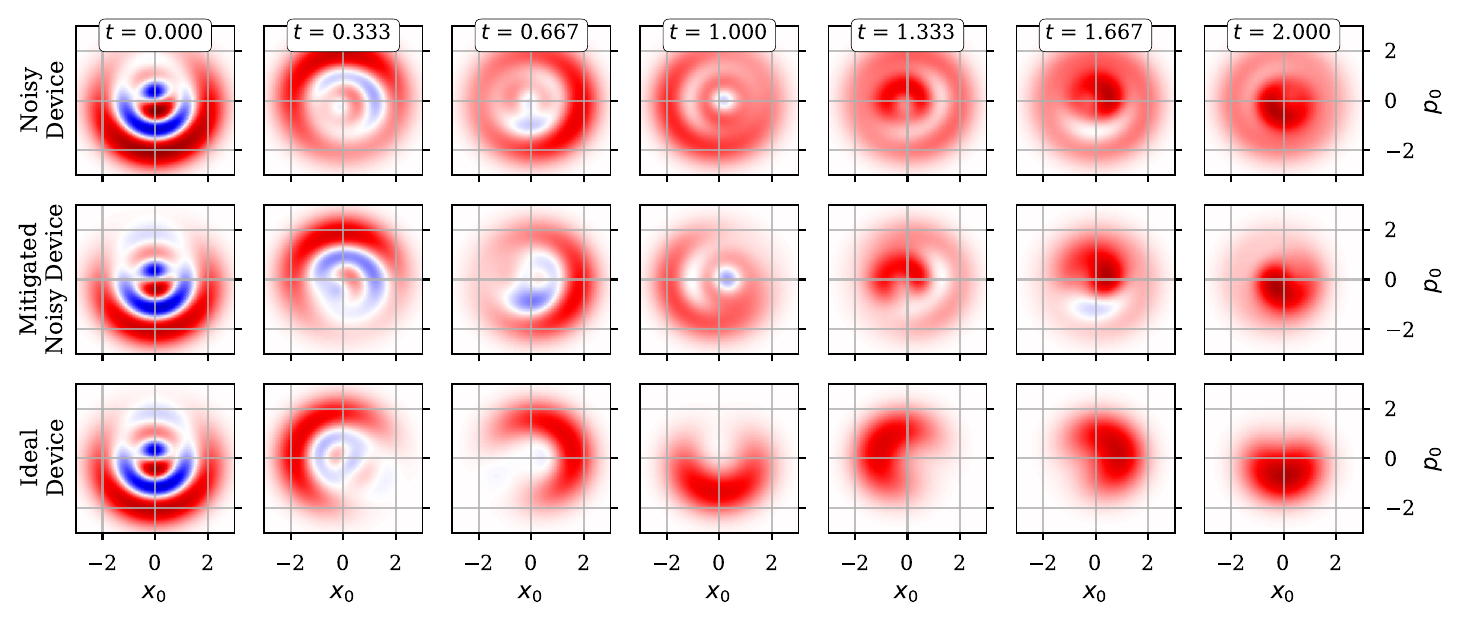}
    \caption{Estimated Wigner quasi-probability distribution of the harmonic oscillator at selected points $t$ along the trajectory. For comparison, the results for the raw IonQ Harmony device (first row), the error-mitigated IonQ Harmony device (second row), and an ideal noiseless simulator (third row) are shown. The red shaded regions indicate where $W > 0$, while the blue shaded regions indicate the regions where $W < 0$. These regions where $W$ is negative correspond to phase space configurations of the system which cannot be explained under a purely classical treatment of the system. }
    \label{fig:qho_ionq_wigner_trajectories}
\end{figure*}

For a harmonic oscillator with density matrix $\rho_{n,n'}$ (where $n,n'$ are eigenstate indices), the Wigner quasi-probability distribution can be computed in closed form using the upper-triangular entries of $\rho$ \cite{bartlett_exact_1949}. This is achieved with the expansion
\begin{widetext}
\begin{equation}
    W(x_0, p_0) = \frac{e^{-(x_0^2 + p_0^2)}}{\pi} \sum_{n}\sum_{n' \geq n} c_{n,n'} \Re\lbrace \rho_{n,n'} [ \sqrt{2}(x_0 + ip_0)]^{n'-n}\rbrace\tilde{L}_n^{(n'-n)}(x_0^2 + p_0^2),
\end{equation}
\end{widetext}
where $\tilde{L}_n^{(k)}$ are the generalized Laguerre polynomials and $c_{n,n'} = (-1)^n (2 - \delta_{n,n'}) \sqrt{n!/n'!}$. In \figref{\ref{fig:qho_ionq_wigner_trajectories}}, we plot the Wigner quasi-probability distributions reconstructed from the density matrices obtained on the IonQ Harmony device (both with and without mitigation) and an ideal device. On the ideal device, we observe that the negative regions of the Wigner distribution (shown in blue) decay rapidly, vanishing at roughly $t \approx 0.6$, yet these regions are present in the noisy device results as late as $t \approx 1.6$. This suggests that applying mitigation to account for device noise may effectively blur the quantum-to-classical transition of the system. From comparing the mitigated and ideal Wigner distributions, we also note the existence of phase mismatching with respect to the ideal distribution, meaning the regions of high quasi-probability mass  are rotated either clockwise or counter-clockwise relative to the ideal distribution. This same phase mismatching is observed in \figref{\ref{fig:qho_ionq_expvals}}, and can be partially attributed to the Harmony device's less precise execution of arbitrary-angle phase gates needed to apply the effective Hamiltonian evolution $T(t)$.

\subsubsection{Noisy Cat State}

In this section, we briefly consider the simulation of a bosonic ``Schrodinger cat" harmonic oscillator state \cite{cochrane_macroscopically_1999}. We give special consideration to the simulation of this state due to the fact that it exhibits parity symmetry that is preserved under the evolution of the system with damping. Due to this property, cat states have been proposed as a platform for realizing discrete qubit states in superconducting and quantum optical systems, and they are instrumental to the implementation of quantum bosonic codes for fault-tolerant computation \cite{michael_new_2016}. Here, we examine the dynamics of the same two-qubit truncated harmonic oscillator representing a single bosonic mode. This mode is prepared in the approximate two-qubit odd cat state
\begin{equation}
        \ket{\psi(0)} = \frac{1}{\sqrt{|\alpha|^2 + |\alpha|^6/3!}}\left(\frac{\alpha^1}{\sqrt{1!}}\ket{1} + \frac{\alpha^3}{\sqrt{3!}}\ket{3} \right).
        \label{eqn:cat_state}
\end{equation}
where $\alpha \in \mathbb{C}$ is a coherent state number. This state is invariant under the parity operator $\hat{\tau} = \sum_{n} (-1)^n\ket{n}\bra{n}$, and remains invariant when evolved with the Lindblad operator $L_1 = \hat{a}$. In the context of bosonic systems, the Lindblad operator $L_1 = \hat{a}$ can be applied to model bosonic particle loss with decay rate $\gamma_1$. 

As the system is simulated on a quantum device, both device noise and sampling error can result in this symmetry being violated. By incorporating this symmetry directly into error mitigation protocols, however, we can impose this symmetry upon the mitigated result, producing results with much higher fidelity than if symmetry is ignored. For example, to impose parity symmetry on the density matrix, we can apply the twirling operation of the group generated by the parity operator $\hat{\tau}$, given by
\begin{equation}
    \mathcal{T}_{\hat{\tau}}(\rho) = \frac{1}{2}(\rho + \hat{\tau}\rho\hat{\tau}^{\dagger}).
    \label{eqn:parity_twirling}
\end{equation}
After the $\mathcal{T}_{\hat{\tau}}$ operation is applied to symmetrize the density matrix, error-mitigation schemes such as QDC inversion can be applied to the system to correct for additional device noise.

\begin{figure}
    \centering
    \begin{tabular}{p{0.95\linewidth}}
        \subfigimg[width=\linewidth]{(a)}{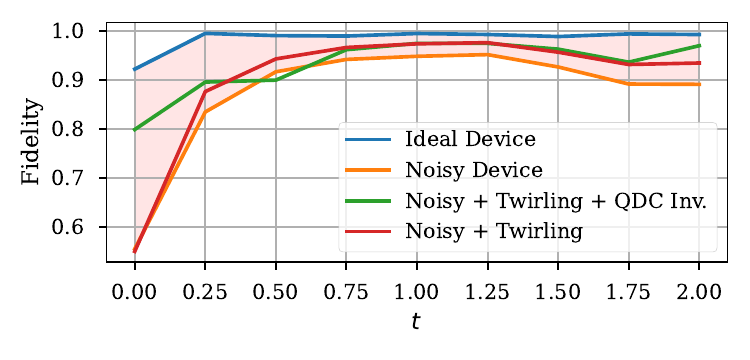} \\
        \subfigimg[width=\linewidth]{(b)}{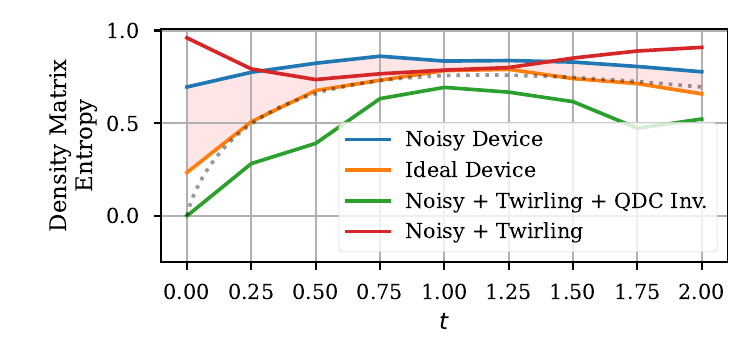} 
    \end{tabular}
    \caption{Fidelity (a) and von Neumann Entropy (b) of the density matrix trajectory obtained through a noise model emulation of the Quantinuum H1-1 device. Results are shown for an ideal noiseless device and the simulated noisy device with no mitigation, with twirling, and with combined twirling and error mitigation. Error mitigation was applied using parity twirling and QDC inversion with a value of $\lambda_{\text{est}} = 0.48$, which was obtained by fitting a QDC to the first state. The ideal trajectory of the system was obtained through a noiseless simulator with the same number of measurements as the noisy device.}
    \label{fig:qho_quantinuum_mitigation}
\end{figure}

To provide a more concrete overview of how this can be incorporated into the error mitigation methods presented in the previous section, we present results obtained from a noisy emulation of the Quantinuum H1-1 device for the initial state in Eq. \eqref{eqn:cat_state} with $\alpha = 1.2$ and $\gamma_1 = 0.6$. We simulate this system using the Kraus series method for nine uniform time steps from $t = 0.0$ to $t = 2.0$. In \figref{\ref{fig:qho_quantinuum_mitigation}} we plot the fidelity and entropy of the cat state trajectory estimated from the noisy device emulation and consider the effects of applying the twirling operation in Eq. \eqref{eqn:parity_twirling} and subsequent QDC inverse error mitigation. From these results, we note that twirling alone provides only a modest increase in overall state fidelity; however, when it is combined with the QDC error mitigation scheme, it provides a significant boost to the overall state fidelity, especially for the initial prepared state at $t = 0$.

In \figref{\ref{fig:qho_ionq_density_trajectories}} we plot the position and momentum densities reconstructed from the noisy simulation results. In the context of bosonic systems, we note that the position and momentum density can be interpreted in the context of the over-complete basis of coherent state numbers $\alpha \in \mathbb{C}$, where $x_0 = \Re[\alpha]$ and $p_0 = \Im[\alpha]$. In the raw noisy position and momentum data (first row) we observe a significant bias that is present in the regions where $x_0 > 0$ and $p_0 > 0$, which is consistent with a strong phase bias in the system qubits, resulting in the appearance of oscillations when $x_0, p_0 < 0$, and no oscillations when $x_0, p_0 > 0$. Since we expect the trajectory to exhibit parity symmetry, we observe that the application of twirling and device noise mitigation produces a more faithful representation of the system dynamics.

\begin{figure*}
    \centering
    \includegraphics[width=0.9\linewidth]{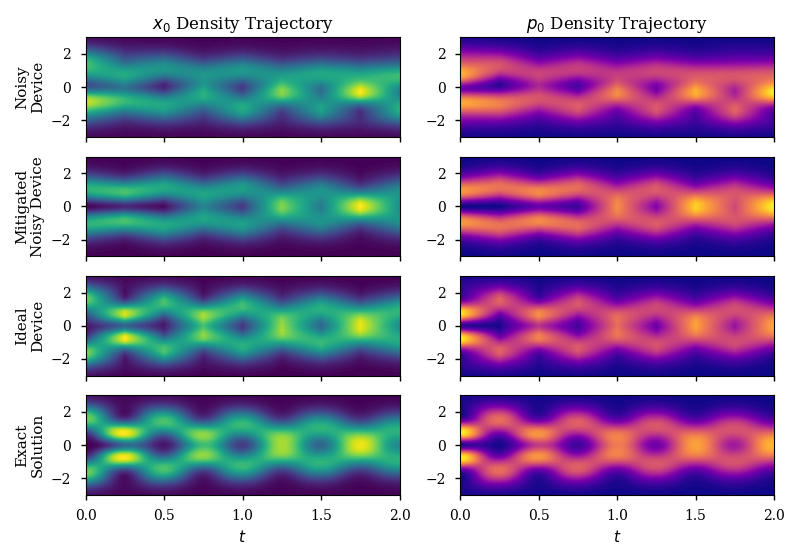}
    \caption{Trajectory of the position density (left column) and momentum density (right column) of the harmonic oscillator prepared in an approximate two-qubit cat state. Densities are shown for a noisy emulated Quantinuum H1-1 device trajectory, the error-mitigated noisy trajectory, an ideal quantum device trajectory, and finally the exact solution to the Lindblad equation.}
    \label{fig:qho_ionq_density_trajectories}
\end{figure*}

\begin{figure*}
    \centering
    \includegraphics[width=\linewidth]{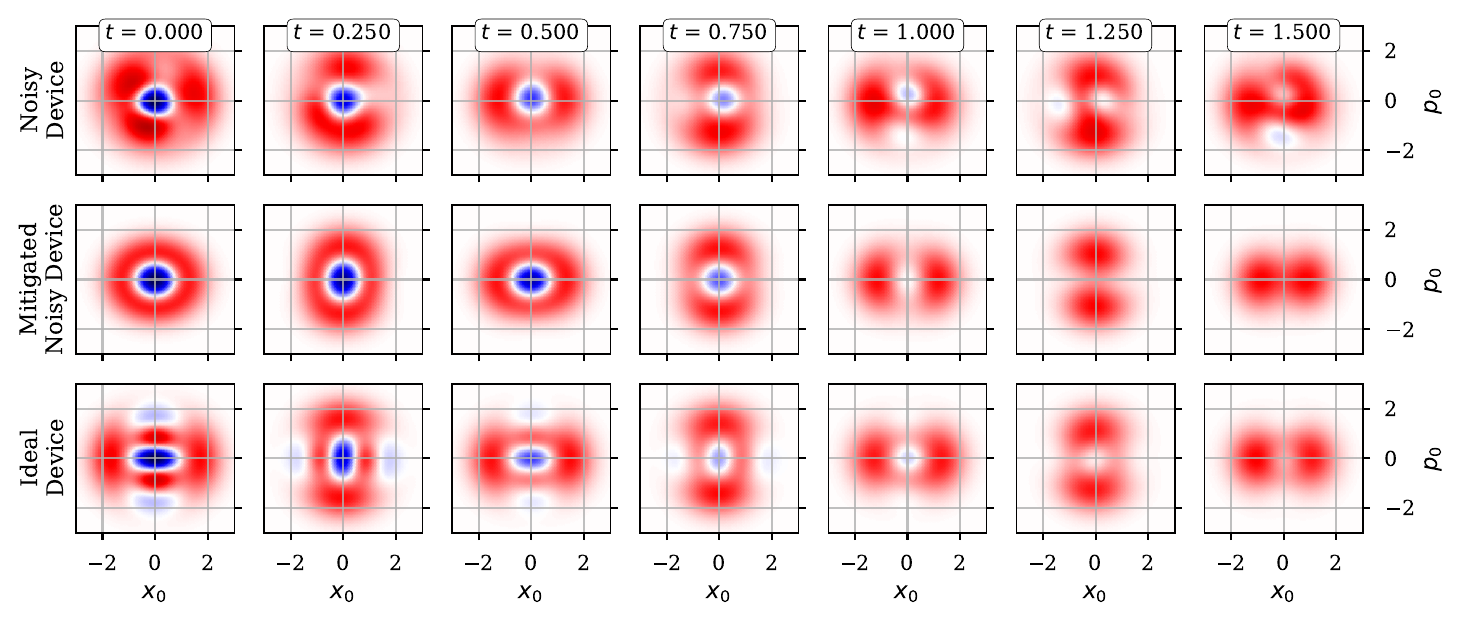}
    \caption{Wigner quasi-probability distribution of the harmonic oscillator modeling a single bosonic mode at selected points $t$ along the trajectory. Results are shown for the emulated Quantinuum H1-1 device (first row), the error-mitigated Quantinuum H1-1 device (second row), and an ideal noiseless simulator (third row). The red shaded regions indicate where $W > 0$, while the blue shaded regions indicate the regions where $W < 0$. These regions where $W$ is negative correspond to phase space configurations of the system which cannot be explained under a purely classical treatment of the system.}
    \label{fig:qho_quantinuum_wigher_trajectories}
\end{figure*}

In \figref{\ref{fig:qho_quantinuum_wigher_trajectories}} we plot the estimated Wigner quasi-probability distributions of the emulated noisy device trajectory, and we compare these distributions to those of an ideal device. In \figref{\ref{fig:qho_quantinuum_mitigation}}), we observed that even the initial state at $t = 0$ was reproduced with low fidelity. From inspecting the Wigner distributions, we observe that this low fidelity partially manifests as broken parity symmetry. When parity twirling is applied to a harmonic oscillator state, the Wigner distribution of that state is averaged with its own $180^\circ$ rotation. It follows that an oscillator state exhibits parity symmetry if its Wigner distribution is invariant under $180^\circ$ rotations. We observe that this is the case for both the mitigated and ideal Wigner distributions, however the mitigated results at around $t = 0$ exhibit distributions that appear to be more consistent with a single excited quantum state (i.e. $\rho \approx \ket{1}\bra{1}$) than with the approximate cat state in Eq. \eqref{eqn:cat_state}. The Wigner distributions of cat states are characterized by the interference pattern of alternating positive and negative regions in phase space which separate two regions of high quasi-probability mass, which are not discernible in the mitigated results for $t < 1.0$.  Although we observe these deviations when the system is in the quantum mechanical regime at $t < 1.0$, we also see that in the classical statistical regime at times $t > 1.0$, there is strong agreement between the mitigated results and the ideal results. This suggests that parity twirling and subsequent error mitigation is very effective when applied to noisy simulations of systems with more classical character, but may not be able to restore all of the important quantum-mechanical effects in systems with a strong quantum character.

\section{Conclusion}\label{sec:conclusion}
We introduced novel methods for efficiently simulating open quantum systems on NISQ devices, to overcome the challenges posed by noise and limited qubit resources. Our approach used Kraus operators to represent the time evolution of open quantum systems. Unlike traditional Trotter-based methods, which can be computationally expensive and noisy due to the large number of quantum gates needed, our method allowed for efficient simulation by avoiding costly Trotterization for certain classes of systems. This is particularly useful for systems that satisfy the commutation relations (i)-(iv), enabling the simulation to be carried out in closed form without requiring diagonalization. We extended the Kraus operator method by deriving a time-perturbative Kraus series that can be applied to simulate the system's evolution under various conditions, such as systems with unitary Lindblad operators, bosonic Lindblad operators, and systems where the Lindblad operators exhibit finite group or semigroup structure. The series was designed to converge rapidly (especially for bosonic systems), making it practical for NISQ devices. We proposed an efficient way to map the Kraus operator series to quantum circuits, making it feasible to simulate these systems on NISQ hardware. The mapping reduced quantum circuit depth and gate count, directly addressing the limitations of NISQ devices in terms of noise and qubit fidelity. We also presented novel error mitigation strategies tailored for open quantum systems simulated on NISQ devices. By fitting noise models such as Pauli and quantum depolarizing channels and applying error mitigation techniques, we demonstrated a significant improvement in the accuracy of the simulations. This approach is particularly relevant for large systems where noise significantly degrades the fidelity of results. Finally, we simulated open quantum systems on real NISQ devices (IonQ Harmony and Quantinuum H1-1), showcasing the practicality and effectiveness of our methods. The novel use of these error mitigation techniques with actual quantum hardware is a critical step in bridging theoretical methods and real-world quantum computing applications. The combination of Kraus-based methods with efficient circuit mapping and error mitigation strategies represents a significant step forward in the simulation of open quantum systems on NISQ devices. 

Our work opens up several avenues for future research. As quantum hardware continues to improve, the methods developed here can be further refined and scaled to tackle more complex quantum systems with larger qubit counts and deeper circuits. An important direction is the extension of these techniques to more diverse classes of open quantum systems, including non-Markovian dynamics, where system-environment correlations persist over time \cite{Head-Marsden_2021,guo2024quantumsimulationopenquantum}. Developing tailored algorithms for these scenarios could unlock more realistic modeling of quantum processes in chemistry, biology, and materials science. 

Furthermore, enhancing the synergy between quantum hardware and software through hardware-aware optimizations is crucial. Circuit depth remains a limitation, but advances in error correction, gate synthesis, and qubit connectivity could allow for more efficient implementations of the Kraus method on future quantum architectures. Collaborative efforts between quantum hardware developers and algorithm designers will be key to pushing the boundaries of quantum simulations on NISQ devices. 

As quantum error mitigation techniques continue to evolve, integrating them with dynamic learning algorithms or hybrid quantum-classical approaches may further improve the reliability of quantum simulations. As we approach fault-tolerant quantum computing, these methods could evolve into robust tools for simulating highly entangled, dissipative systems that are beyond the reach of classical computation. 

\subsection*{Data Availability Statement}

The code, data, and demo notebooks necessary to reproduce these results can be found in the associated Github repository \cite{burdine_2024}.

\begin{acknowledgments}
N. B. and G. S. acknowledge 
support by the National Science Foundation under award DGE-2152168 and the Department of Energy under awards DE-SC0024325 and DE-SC0024328. This research used resources of the Oak Ridge Leadership Computing Facility, which is a DOE Office of Science User Facility supported under Contract DE-AC05-00OR22725. This research was also supported by compute credits from IonQ via the Microsoft Azure Quantum Credits program.
\end{acknowledgments}

\subsection*{Author Contributions}
C.B. and E.P.B conceptualized the work. C.B. developed the theory and code. C.B., N.B., and G.S. conducted the quantum hardware experiments. All authors contributed to the writing and review of the manuscript.
\appendix


\section{Mapping the Kraus Series to Quantum Circuits}
\label{sec:appendix_circuits}

In this appendix, we give additional details concerning the mapping of Kraus operators to quantum circuits and provide some examples of how this can be done in an efficient manner for systems satisfying the commutation relations (i)-(iv). In \secref{\ref{sec:background_mapping_to_circuits}}, we discussed briefly how these Kraus operators can be implemented using unitary dilations of the Lindblad operators $U_{L_n}$ and a unitary dilation of the effective time-evolution operator $T(t)$. We also discussed in \secref{\ref{sec:background_special_cases}} how certain special cases (such as bosonic systems and systems that form finite abelian groups) can be simulated with greater efficiency.

First, we consider the problem of realizing the effective evolution operator $T(t)$. As discussed in \secref{\ref{sec:background_mapping_to_circuits}}, the commutation relations (i) and (ii) ensure that the effective Hamiltonian $H_{\text{eff}}$ (given in Eq. \eqref{eqn:H_eff}) can be diagonalized in the same basis as the system Hamiltonian $H$. It follows that $T(t)$ can then be diagonalized in the form
\begin{equation}
    T(t) = U W(t)\Lambda(T) U^{\dagger}
\end{equation}

where $W(t)$ is diagonal unitary and $\Lambda(t)$ is diagonal positive semi-definite with time-dependent entries of the form $W_{ii}(t) = e^{-i\theta_i t}$ and $\Lambda_{ii}(t) = e^{-\beta_i t}$, where $\theta_i \in [0,2\pi)$ and $\beta_i \geq 0$ correspond to the imaginary and real components of $-it E_{\text{eff}}/\hbar$, where $E_{\text{eff}}$ (given in Eq. \eqref{eqn:E_eff_eigenvalues}) is an eigenvalue of $H_{\text{eff}}$. While one could in theory apply the diagonalizing unitary $U$ in the quantum circuit, this is often impractical on NISQ devices, since one might as well represent the entire system in this diagonal basis and avoid the expense of applying $U$ altogether.

Methods of efficiently implementing diagonal unitaries $W(t)$ have been discussed in the literature (e.g., see \cite{bullock_smaller_2003}). Among these methods, there exists a general trade-off between classical pre-processing complexity and quantum gate complexity, depending on whether the diagonal elements are encoded with parameterized rotation gates via a standard binary or a Gray code encoding. In a binary encoding, the elements $W_{ii}(t)$ of a $d$-dimensional unitary are encoded through a sequence of $d$ multi-qubit controlled phase gates, where the control qubits for the $n$th phase gate are the $1$ bits in the binary representation of $n$. In this encoding, the phase angles corresponding to the entries $W_{ii}(t)$ can be computed efficiently in time $O(d\log(d))$ and can be executed with $O(d\log(d))$ quantum gates \cite{burdine_trotterless_2024}. In a Gray code encoding, controlled rotation gates are applied to single qubits such that a sequence of parity-controlled $R_z$ gates are applied to recursive sub-blocks of $W$ \cite{bullock_smaller_2003}. An important benefit of the Gray code encoding is that it requires roughly $d/2$ CNOT gates when $d$ is a power of $2$; however, the mapping of the entries $W_ii(t)$ to rotation gate parameters requires solving a $d^2 \times d^2$ linear system, which has a classical complexity between $O(d^2)$ and $O(d^3)$ that dominates the quantum simulation time for large systems. For this reason, we consider the standard binary encoding more favorable when simulating the trajectory of large systems over many $t$ values, and the Gray code encoding more favorable for small systems with few $t$ values. In the results presented in \secref{\ref{sec:results}}. 
The non-unitary diagonal matrix $\Lambda(t)$ can also be mapped to quantum gates through methods similar to those for the unitary $W(t)$; however instead of applying phase or $R_z$ gates, controlled $R_y$ gates are instead applied on an ancilla qubit with the system qubit as a control. This implements a unitary block encoding of the real diagonal operator $\Lambda(t)$. Both the standard binary encoding and the Gray code encoding of this non-unitary operator exhibit similar asymptotic quantum gate complexity and classical parameter mapping complexity as the diagonal opertor $W(t)$. For additional details regarding the trade-off, and for circuit diagrams implementing $\Lambda(t)$ and $W(t)$ for simple systems (such as the damped quantum harmonic oscillator), we refer the reader to reference \cite{burdine_trotterless_2024}.

\section{Pauli Noise Models on NISQ Hardware}
\label{sec:appendix_pauli_noise_models}

In this appendix, we consider Pauli noise models, and how they can be applied to density matrix simulations of open quantum systems on NISQ hardware. In \secref{\ref{sec:qho_error_mitigation}}, we introduced the general Pauli channel noise model Eq. \eqref{eqn:pauli_channel_model}, which represents the Markovian evolution of a system under a continuous-time Pauli channel for some unknown time. We also considered a special case of the Pauli channel model, called the quantum depolarizing channel (QDC), given by Eq. \eqref{eqn:depolarizing_channel_model}. Although it is a relatively naive noise model, a QDC can be applied to model depolarization of quantum states, which is a type of decoherence that commutes with all other types of decoherence described by a CPTP map. Specifically, we claimed that the QDC satisfies the commutativity property Eq. \eqref{eqn:depolarizing_channel_commutativity}, where $\mathcal{E}_{\text{sim}}$ is the CPTP map being simulated on the quantum device. This can be shown by expanding $\mathcal{E}_{\text{sim}}$ as a sum of Kraus operators $K_i$, so that by the unitality of the Kraus operators we obtain
\begin{equation}
    \begin{aligned}
        \mathcal{E}_{\text{sim}}(\mathcal{E}_{\text{QDC}}(\rho)) 
        &=\sum_i K_i \left[(1-\lambda)\rho + \frac{\lambda}{2^N}I \right]K_i^{\dagger} \\
        &= \sum_i (1-\lambda) K_i\rho K_i^{\dagger} + \frac{\lambda}{2^N} K_iK_i^{\dagger} \\
        &= (1-\lambda)\mathcal{E}_{\text{sim}}(\rho) + \frac{\lambda}{2^N}I \\
        &= \mathcal{E}_{\text{QDC}}(\mathcal{E}_{\text{sim}}(\rho))
    \end{aligned}
\end{equation}
as claimed. In a similar manner, it can also be shown that any two Pauli channels commute. Letting $P_0 \equiv I$ for convenience of notation, we see that
\begin{equation}
    \begin{aligned}
        \mathcal{E}_{\text{PC}}(\mathcal{E}_{\text{PC}}'(\rho))
        &= \sum_{i} \varepsilon_i P_i \left[ \sum_j \varepsilon_j' P_j' \rho P_j'^{\dagger} \right] P_i^{\dagger} \\
        &= \sum_{i}\sum_j \varepsilon_i\varepsilon_j' P_i P_j' \rho P_j'^{\dagger} P_i^{\dagger} \\
        &= \sum_{j}\sum_i \varepsilon_j'\varepsilon_i P_j' P_i \rho P_i^{\dagger} P_j'^{\dagger} \\
        &= \mathcal{E}_{\text{PC}}'(\mathcal{E}_{\text{PC}}(\rho))
    \end{aligned}
\end{equation}
as claimed. This result should come as no surprise, because we have already shown that the Lindblad operators $L_n$ of a continuous-time Pauli channel form a commutative group, hence the CPTP maps they generate should also commute.

Next, we consider the problem of fitting and inverting Pauli channels to measured data. Given a set of ``exact" density matrices $\rho_1, \rho_2, ..., \rho_n$ and ``noisy" density matrices $\tilde{\rho}_1, \tilde{\rho}_2, ..., \tilde{\rho}_n$, we aim to find error parameters $\varepsilon_i$ such that the mean Frobenius norm error $\sum_{i=1}^n \norm{\mathcal{E}_{PC}(\rho_i) - \tilde{\rho}_i}_F$ for all noisy density matrices is minimized. This corresponds to the non-negative least squares regression problem
\begin{equation}
    \begin{aligned}
    &\text{Minimize:} \ \frac{1}{n}\sum_{j=1}^{n} \left\lVert \tilde{\rho}_j - \sum_{i} \varepsilon_i P_i \rho_j P_i^{\dagger}\right\rVert_F^2\  \\[1mm]
    &\text{subject to:} \ \varepsilon_i \geq 0
    \end{aligned}
    \label{eqn:nnls_problem}
\end{equation}
which can be solved through standard quadratic programming algorithms. A similar method can be used for the QDC. An important feature of Pauli noise models is that they can be inverted. Specifically, the action of a channel $\mathcal{E}(\rho)$ can be written as a matrix $A$ acting on the vectorized density matrix $\vec{\rho}$:
\begin{equation}
    A_{\mathcal{E}}\vec{\rho} = \vec{\tilde{\rho}}
\end{equation}
where $A$ is the $d^2 \times d^2$ matrix with entries
\begin{equation}
    A_{\mathcal{E}} = \sum_i \varepsilon_i (P_i \otimes \overline{P_i})
\end{equation}
where $\overline{P_i}$ denotes the complex conjugate of $P_i$. 

Given a noisy estimate of a density matrix $\tilde{\rho}$, one can compute the mitigated density matrix $\rho_{\text{mit}}$ by applying the channel inverse. This is computed as
\begin{equation}
    \vec{\rho}_{\text{mit}} = A_{\mathcal{E}}^{+}\vec{\tilde{\rho}}.
\end{equation}
where $A^{+}$ denotes the matrix pseudoinverse. In the case of the QDC, the inverse can be computed with the much simpler formula
\begin{equation}
    \rho_{\text{mit}} = \frac{1}{(1-\lambda)}(\tilde{\rho} - (\lambda/2^N)I).
\end{equation}
For both the QDC and the general Pauli channel, applying the channel inverse produces a density matrix $\rho_{\text{mit}}$ that is Hermitian with unit trace. Sometimes, however, the resulting density matrices have small negative eigenvalues. These ``unphysical" density matrices can be corrected by fixing all negative eigenvalues to zero and re-normalizing $\rho_{\text{mit}}$ to have unit trace.

Because computing the QDC inverse is computationally inexpensive, we also note that the parameter $\lambda$ can be estimated based on minimizing the reconstruction error of $\rho_{\text{mit}}$ versus a known state without device noise. This can be done by either minimizing the mean Frobenius error $\sum_{j=1}^n \norm{\rho_{j} - \rho_{\text{mit}j}}/ n$ or by maximizing the fidelity of the reconstructed density matrix, which is computed as:
\begin{equation}
    F(\rho,\rho_{\text{mit}}) = \left(\Tr\left[(\rho^{\frac{1}{2}} \rho_{\text{mit}} \rho^{\frac{1}{2}})^{\frac{1}{2}}\right]\right)^2
\label{eqn:fidelity}
\end{equation}

In this paper, we maximize Eq. \eqref{eqn:fidelity} to find the optimal estimated value of $\lambda$ for the QDC as shown in \figref{\ref{fig:qho_ionq_qdc}(b)}. For the general Pauli channel, however, we use the non-negative least squares approach that solves Eq. \eqref{eqn:nnls_problem} for the optimal coefficients $\varepsilon_i$, as shown in \figref{\ref{fig:qho_ionq_qdc}}.

\bibliography{main}

\begin{thebibliography}{55}%
\makeatletter
\providecommand \@ifxundefined [1]{%
 \@ifx{#1\undefined}
}%
\providecommand \@ifnum [1]{%
 \ifnum #1\expandafter \@firstoftwo
 \else \expandafter \@secondoftwo
 \fi
}%
\providecommand \@ifx [1]{%
 \ifx #1\expandafter \@firstoftwo
 \else \expandafter \@secondoftwo
 \fi
}%
\providecommand \natexlab [1]{#1}%
\providecommand \enquote  [1]{``#1''}%
\providecommand \bibnamefont  [1]{#1}%
\providecommand \bibfnamefont [1]{#1}%
\providecommand \citenamefont [1]{#1}%
\providecommand \href@noop [0]{\@secondoftwo}%
\providecommand \href [0]{\begingroup \@sanitize@url \@href}%
\providecommand \@href[1]{\@@startlink{#1}\@@href}%
\providecommand \@@href[1]{\endgroup#1\@@endlink}%
\providecommand \@sanitize@url [0]{\catcode `\\12\catcode `\$12\catcode `\&12\catcode `\#12\catcode `\^12\catcode `\_12\catcode `\%12\relax}%
\providecommand \@@startlink[1]{}%
\providecommand \@@endlink[0]{}%
\providecommand \url  [0]{\begingroup\@sanitize@url \@url }%
\providecommand \@url [1]{\endgroup\@href {#1}{\urlprefix }}%
\providecommand \urlprefix  [0]{URL }%
\providecommand \Eprint [0]{\href }%
\providecommand \doibase [0]{https://doi.org/}%
\providecommand \selectlanguage [0]{\@gobble}%
\providecommand \bibinfo  [0]{\@secondoftwo}%
\providecommand \bibfield  [0]{\@secondoftwo}%
\providecommand \translation [1]{[#1]}%
\providecommand \BibitemOpen [0]{}%
\providecommand \bibitemStop [0]{}%
\providecommand \bibitemNoStop [0]{.\EOS\space}%
\providecommand \EOS [0]{\spacefactor3000\relax}%
\providecommand \BibitemShut  [1]{\csname bibitem#1\endcsname}%
\let\auto@bib@innerbib\@empty
\bibitem [{\citenamefont {Cleve}\ and\ \citenamefont {Wang}(2017)}]{cleve_et_al:LIPIcs.ICALP.2017.17}%
  \BibitemOpen
  \bibfield  {author} {\bibinfo {author} {\bibfnamefont {R.}~\bibnamefont {Cleve}}\ and\ \bibinfo {author} {\bibfnamefont {C.}~\bibnamefont {Wang}},\ }\bibfield  {title} {\bibinfo {title} {{Efficient Quantum Algorithms for Simulating Lindblad Evolution}},\ }in\ \href {https://doi.org/10.4230/LIPIcs.ICALP.2017.17} {\emph {\bibinfo {booktitle} {44th International Colloquium on Automata, Languages, and Programming (ICALP 2017)}}},\ \bibinfo {series} {Leibniz International Proceedings in Informatics (LIPIcs)}, Vol.~\bibinfo {volume} {80},\ \bibinfo {editor} {edited by\ \bibinfo {editor} {\bibfnamefont {I.}~\bibnamefont {Chatzigiannakis}}, \bibinfo {editor} {\bibfnamefont {P.}~\bibnamefont {Indyk}}, \bibinfo {editor} {\bibfnamefont {F.}~\bibnamefont {Kuhn}},\ and\ \bibinfo {editor} {\bibfnamefont {A.}~\bibnamefont {Muscholl}}}\ (\bibinfo  {publisher} {Schloss Dagstuhl -- Leibniz-Zentrum f{\"u}r Informatik},\ \bibinfo {address} {Dagstuhl, Germany},\ \bibinfo {year} {2017})\ pp.\ \bibinfo {pages}
  {17:1--17:14}\BibitemShut {NoStop}%
\bibitem [{\citenamefont {Han}\ \emph {et~al.}(2021)\citenamefont {Han}, \citenamefont {Cai}, \citenamefont {Hu}, \citenamefont {Mu}, \citenamefont {Ma}, \citenamefont {Xu}, \citenamefont {Wang}, \citenamefont {Wang}, \citenamefont {Song}, \citenamefont {Zou},\ and\ \citenamefont {Sun}}]{Han_2021}%
  \BibitemOpen
  \bibfield  {author} {\bibinfo {author} {\bibfnamefont {J.}~\bibnamefont {Han}}, \bibinfo {author} {\bibfnamefont {W.}~\bibnamefont {Cai}}, \bibinfo {author} {\bibfnamefont {L.}~\bibnamefont {Hu}}, \bibinfo {author} {\bibfnamefont {X.}~\bibnamefont {Mu}}, \bibinfo {author} {\bibfnamefont {Y.}~\bibnamefont {Ma}}, \bibinfo {author} {\bibfnamefont {Y.}~\bibnamefont {Xu}}, \bibinfo {author} {\bibfnamefont {W.}~\bibnamefont {Wang}}, \bibinfo {author} {\bibfnamefont {H.}~\bibnamefont {Wang}}, \bibinfo {author} {\bibfnamefont {Y.~P.}\ \bibnamefont {Song}}, \bibinfo {author} {\bibfnamefont {C.-L.}\ \bibnamefont {Zou}},\ and\ \bibinfo {author} {\bibfnamefont {L.}~\bibnamefont {Sun}},\ }\bibfield  {title} {\bibinfo {title} {Experimental simulation of open quantum system dynamics via trotterization},\ }\bibfield  {journal} {\bibinfo  {journal} {Physical Review Letters}\ }\textbf {\bibinfo {volume} {127}},\ \href {https://doi.org/10.1103/physrevlett.127.020504} {10.1103/physrevlett.127.020504} (\bibinfo {year}
  {2021})\BibitemShut {NoStop}%
\bibitem [{\citenamefont {Childs}\ \emph {et~al.}(2021{\natexlab{a}})\citenamefont {Childs}, \citenamefont {Su}, \citenamefont {Tran}, \citenamefont {Wiebe},\ and\ \citenamefont {Zhu}}]{Childs_2021}%
  \BibitemOpen
  \bibfield  {author} {\bibinfo {author} {\bibfnamefont {A.~M.}\ \bibnamefont {Childs}}, \bibinfo {author} {\bibfnamefont {Y.}~\bibnamefont {Su}}, \bibinfo {author} {\bibfnamefont {M.~C.}\ \bibnamefont {Tran}}, \bibinfo {author} {\bibfnamefont {N.}~\bibnamefont {Wiebe}},\ and\ \bibinfo {author} {\bibfnamefont {S.}~\bibnamefont {Zhu}},\ }\bibfield  {title} {\bibinfo {title} {Theory of trotter error with commutator scaling},\ }\href {https://doi.org/10.1103/PhysRevX.11.011020} {\bibfield  {journal} {\bibinfo  {journal} {Phys. Rev. X}\ }\textbf {\bibinfo {volume} {11}},\ \bibinfo {pages} {011020} (\bibinfo {year} {2021}{\natexlab{a}})}\BibitemShut {NoStop}%
\bibitem [{\citenamefont {Chen}\ \emph {et~al.}(2024)\citenamefont {Chen}, \citenamefont {Gomes}, \citenamefont {Niu},\ and\ \citenamefont {Jong}}]{Chen_2024}%
  \BibitemOpen
  \bibfield  {author} {\bibinfo {author} {\bibfnamefont {H.}~\bibnamefont {Chen}}, \bibinfo {author} {\bibfnamefont {N.}~\bibnamefont {Gomes}}, \bibinfo {author} {\bibfnamefont {S.}~\bibnamefont {Niu}},\ and\ \bibinfo {author} {\bibfnamefont {W.~A.~d.}\ \bibnamefont {Jong}},\ }\bibfield  {title} {\bibinfo {title} {Adaptive variational simulation for open quantum systems},\ }\href {https://doi.org/10.22331/q-2024-02-13-1252} {\bibfield  {journal} {\bibinfo  {journal} {Quantum}\ }\textbf {\bibinfo {volume} {8}},\ \bibinfo {pages} {1252} (\bibinfo {year} {2024})}\BibitemShut {NoStop}%
\bibitem [{\citenamefont {Endo}\ \emph {et~al.}(2020)\citenamefont {Endo}, \citenamefont {Sun}, \citenamefont {Li}, \citenamefont {Benjamin},\ and\ \citenamefont {Yuan}}]{Endo_2020}%
  \BibitemOpen
  \bibfield  {author} {\bibinfo {author} {\bibfnamefont {S.}~\bibnamefont {Endo}}, \bibinfo {author} {\bibfnamefont {J.}~\bibnamefont {Sun}}, \bibinfo {author} {\bibfnamefont {Y.}~\bibnamefont {Li}}, \bibinfo {author} {\bibfnamefont {S.~C.}\ \bibnamefont {Benjamin}},\ and\ \bibinfo {author} {\bibfnamefont {X.}~\bibnamefont {Yuan}},\ }\bibfield  {title} {\bibinfo {title} {Variational quantum simulation of general processes},\ }\href {https://doi.org/10.1103/PhysRevLett.125.010501} {\bibfield  {journal} {\bibinfo  {journal} {Phys. Rev. Lett.}\ }\textbf {\bibinfo {volume} {125}},\ \bibinfo {pages} {010501} (\bibinfo {year} {2020})}\BibitemShut {NoStop}%
\bibitem [{\citenamefont {Mellak}\ \emph {et~al.}(2024)\citenamefont {Mellak}, \citenamefont {Arrigoni},\ and\ \citenamefont {von~der Linden}}]{Mellak2024DeepNN}%
  \BibitemOpen
  \bibfield  {author} {\bibinfo {author} {\bibfnamefont {J.}~\bibnamefont {Mellak}}, \bibinfo {author} {\bibfnamefont {E.}~\bibnamefont {Arrigoni}},\ and\ \bibinfo {author} {\bibfnamefont {W.}~\bibnamefont {von~der Linden}},\ }\bibfield  {title} {\bibinfo {title} {Deep neural networks as variational solutions for correlated open quantum systems},\ }\href {https://api.semanticscholar.org/CorpusID:267212083} {\bibfield  {journal} {\bibinfo  {journal} {Communications Physics}\ }\textbf {\bibinfo {volume} {7}},\ \bibinfo {pages} {1} (\bibinfo {year} {2024})}\BibitemShut {NoStop}%
\bibitem [{\citenamefont {Carmichael}(1993)}]{Carmichael_1993}%
  \BibitemOpen
  \bibfield  {author} {\bibinfo {author} {\bibfnamefont {H.~J.}\ \bibnamefont {Carmichael}},\ }\bibfield  {title} {\bibinfo {title} {Quantum trajectory theory for cascaded open systems},\ }\href {https://doi.org/10.1103/PhysRevLett.70.2273} {\bibfield  {journal} {\bibinfo  {journal} {Phys. Rev. Lett.}\ }\textbf {\bibinfo {volume} {70}},\ \bibinfo {pages} {2273} (\bibinfo {year} {1993})}\BibitemShut {NoStop}%
\bibitem [{\citenamefont {Cech}\ \emph {et~al.}(2023)\citenamefont {Cech}, \citenamefont {Lesanovsky},\ and\ \citenamefont {Carollo}}]{Cech_2023}%
  \BibitemOpen
  \bibfield  {author} {\bibinfo {author} {\bibfnamefont {M.}~\bibnamefont {Cech}}, \bibinfo {author} {\bibfnamefont {I.}~\bibnamefont {Lesanovsky}},\ and\ \bibinfo {author} {\bibfnamefont {F.}~\bibnamefont {Carollo}},\ }\bibfield  {title} {\bibinfo {title} {Thermodynamics of quantum trajectories on a quantum computer},\ }\href {https://doi.org/10.1103/PhysRevLett.131.120401} {\bibfield  {journal} {\bibinfo  {journal} {Phys. Rev. Lett.}\ }\textbf {\bibinfo {volume} {131}},\ \bibinfo {pages} {120401} (\bibinfo {year} {2023})}\BibitemShut {NoStop}%
\bibitem [{\citenamefont {Guimar\~aes}\ \emph {et~al.}(2023)\citenamefont {Guimar\~aes}, \citenamefont {Lim}, \citenamefont {Vasilevskiy}, \citenamefont {Huelga},\ and\ \citenamefont {Plenio}}]{Guimaraes_2023}%
  \BibitemOpen
  \bibfield  {author} {\bibinfo {author} {\bibfnamefont {J.~D.}\ \bibnamefont {Guimar\~aes}}, \bibinfo {author} {\bibfnamefont {J.}~\bibnamefont {Lim}}, \bibinfo {author} {\bibfnamefont {M.~I.}\ \bibnamefont {Vasilevskiy}}, \bibinfo {author} {\bibfnamefont {S.~F.}\ \bibnamefont {Huelga}},\ and\ \bibinfo {author} {\bibfnamefont {M.~B.}\ \bibnamefont {Plenio}},\ }\bibfield  {title} {\bibinfo {title} {Noise-assisted digital quantum simulation of open systems using partial probabilistic error cancellation},\ }\href {https://doi.org/10.1103/PRXQuantum.4.040329} {\bibfield  {journal} {\bibinfo  {journal} {PRX Quantum}\ }\textbf {\bibinfo {volume} {4}},\ \bibinfo {pages} {040329} (\bibinfo {year} {2023})}\BibitemShut {NoStop}%
\bibitem [{\citenamefont {Cygorek}\ \emph {et~al.}(2021)\citenamefont {Cygorek}, \citenamefont {Cosacchi}, \citenamefont {Vagov}, \citenamefont {Axt}, \citenamefont {Lovett}, \citenamefont {Keeling},\ and\ \citenamefont {Gauger}}]{Cygorek2021SimulationOO}%
  \BibitemOpen
  \bibfield  {author} {\bibinfo {author} {\bibfnamefont {M.}~\bibnamefont {Cygorek}}, \bibinfo {author} {\bibfnamefont {M.}~\bibnamefont {Cosacchi}}, \bibinfo {author} {\bibfnamefont {A.}~\bibnamefont {Vagov}}, \bibinfo {author} {\bibfnamefont {V.~M.}\ \bibnamefont {Axt}}, \bibinfo {author} {\bibfnamefont {B.~W.}\ \bibnamefont {Lovett}}, \bibinfo {author} {\bibfnamefont {J.}~\bibnamefont {Keeling}},\ and\ \bibinfo {author} {\bibfnamefont {E.~M.}\ \bibnamefont {Gauger}},\ }\bibfield  {title} {\bibinfo {title} {Simulation of open quantum systems by automated compression of arbitrary environments},\ }\href {https://api.semanticscholar.org/CorpusID:247724725} {\bibfield  {journal} {\bibinfo  {journal} {Nature Physics}\ }\textbf {\bibinfo {volume} {18}},\ \bibinfo {pages} {662 } (\bibinfo {year} {2021})}\BibitemShut {NoStop}%
\bibitem [{\citenamefont {Kamakari}\ \emph {et~al.}(2022)\citenamefont {Kamakari}, \citenamefont {Sun}, \citenamefont {Motta},\ and\ \citenamefont {Minnich}}]{Kamakari_2022}%
  \BibitemOpen
  \bibfield  {author} {\bibinfo {author} {\bibfnamefont {H.}~\bibnamefont {Kamakari}}, \bibinfo {author} {\bibfnamefont {S.-N.}\ \bibnamefont {Sun}}, \bibinfo {author} {\bibfnamefont {M.}~\bibnamefont {Motta}},\ and\ \bibinfo {author} {\bibfnamefont {A.~J.}\ \bibnamefont {Minnich}},\ }\bibfield  {title} {\bibinfo {title} {Digital quantum simulation of open quantum systems using quantum imaginary--time evolution},\ }\href {https://doi.org/10.1103/PRXQuantum.3.010320} {\bibfield  {journal} {\bibinfo  {journal} {PRX Quantum}\ }\textbf {\bibinfo {volume} {3}},\ \bibinfo {pages} {010320} (\bibinfo {year} {2022})}\BibitemShut {NoStop}%
\bibitem [{\citenamefont {Hu}\ \emph {et~al.}(2019)\citenamefont {Hu}, \citenamefont {Xia},\ and\ \citenamefont {Kais}}]{Hu2019AQA}%
  \BibitemOpen
  \bibfield  {author} {\bibinfo {author} {\bibfnamefont {Z.}~\bibnamefont {Hu}}, \bibinfo {author} {\bibfnamefont {R.}~\bibnamefont {Xia}},\ and\ \bibinfo {author} {\bibfnamefont {S.}~\bibnamefont {Kais}},\ }\bibfield  {title} {\bibinfo {title} {A quantum algorithm for evolving open quantum dynamics on quantum computing devices},\ }\href {https://api.semanticscholar.org/CorpusID:90234318} {\bibfield  {journal} {\bibinfo  {journal} {Scientific Reports}\ }\textbf {\bibinfo {volume} {10}} (\bibinfo {year} {2019})}\BibitemShut {NoStop}%
\bibitem [{\citenamefont {Kim}\ \emph {et~al.}(2022)\citenamefont {Kim}, \citenamefont {Nichol}, \citenamefont {Jordan},\ and\ \citenamefont {Franco}}]{Kim_2022}%
  \BibitemOpen
  \bibfield  {author} {\bibinfo {author} {\bibfnamefont {C.~W.}\ \bibnamefont {Kim}}, \bibinfo {author} {\bibfnamefont {J.~M.}\ \bibnamefont {Nichol}}, \bibinfo {author} {\bibfnamefont {A.~N.}\ \bibnamefont {Jordan}},\ and\ \bibinfo {author} {\bibfnamefont {I.}~\bibnamefont {Franco}},\ }\bibfield  {title} {\bibinfo {title} {Analog quantum simulation of the dynamics of open quantum systems with quantum dots and microelectronic circuits},\ }\href {https://doi.org/10.1103/PRXQuantum.3.040308} {\bibfield  {journal} {\bibinfo  {journal} {PRX Quantum}\ }\textbf {\bibinfo {volume} {3}},\ \bibinfo {pages} {040308} (\bibinfo {year} {2022})}\BibitemShut {NoStop}%
\bibitem [{\citenamefont {Breuer}\ and\ \citenamefont {Petruccione}(2002)}]{breuer_theory_2002}%
  \BibitemOpen
  \bibfield  {author} {\bibinfo {author} {\bibfnamefont {H.-P.}\ \bibnamefont {Breuer}}\ and\ \bibinfo {author} {\bibfnamefont {F.}~\bibnamefont {Petruccione}},\ }\href@noop {} {\emph {\bibinfo {title} {The {Theory} of {Open} {Quantum} {Systems}}}}\ (\bibinfo  {publisher} {Oxford University Press},\ \bibinfo {year} {2002})\BibitemShut {NoStop}%
\bibitem [{\citenamefont {Trotter}(1959)}]{trotter_product_1959}%
  \BibitemOpen
  \bibfield  {author} {\bibinfo {author} {\bibfnamefont {H.~F.}\ \bibnamefont {Trotter}},\ }\bibfield  {title} {\bibinfo {title} {On the product of semi-groups of operators},\ }\href {https://doi.org/10.1090/S0002-9939-1959-0108732-6} {\bibfield  {journal} {\bibinfo  {journal} {Proceedings of the American Mathematical Society}\ }\textbf {\bibinfo {volume} {10}},\ \bibinfo {pages} {545} (\bibinfo {year} {1959})}\BibitemShut {NoStop}%
\bibitem [{\citenamefont {Hatano}\ and\ \citenamefont {Suzuki}(2005)}]{hatano_finding_2005}%
  \BibitemOpen
  \bibfield  {author} {\bibinfo {author} {\bibfnamefont {N.}~\bibnamefont {Hatano}}\ and\ \bibinfo {author} {\bibfnamefont {M.}~\bibnamefont {Suzuki}},\ }\bibfield  {title} {\bibinfo {title} {Finding {Exponential} {Product} {Formulas} of {Higher} {Orders}},\ }in\ \href {https://doi.org/10.1007/11526216_2} {\emph {\bibinfo {booktitle} {Quantum {Annealing} and {Other} {Optimization} {Methods}}}},\ \bibinfo {series and number} {Lecture {Notes} in {Physics}},\ \bibinfo {editor} {edited by\ \bibinfo {editor} {\bibfnamefont {A.}~\bibnamefont {Das}}\ and\ \bibinfo {editor} {\bibfnamefont {B.}~\bibnamefont {K.~Chakrabarti}}}\ (\bibinfo  {publisher} {Springer},\ \bibinfo {address} {Berlin, Heidelberg},\ \bibinfo {year} {2005})\ pp.\ \bibinfo {pages} {37--68}\BibitemShut {NoStop}%
\bibitem [{\citenamefont {Childs}\ \emph {et~al.}(2021{\natexlab{b}})\citenamefont {Childs}, \citenamefont {Su}, \citenamefont {Tran}, \citenamefont {Wiebe},\ and\ \citenamefont {Zhu}}]{childs_theory_2021}%
  \BibitemOpen
  \bibfield  {author} {\bibinfo {author} {\bibfnamefont {A.~M.}\ \bibnamefont {Childs}}, \bibinfo {author} {\bibfnamefont {Y.}~\bibnamefont {Su}}, \bibinfo {author} {\bibfnamefont {M.~C.}\ \bibnamefont {Tran}}, \bibinfo {author} {\bibfnamefont {N.}~\bibnamefont {Wiebe}},\ and\ \bibinfo {author} {\bibfnamefont {S.}~\bibnamefont {Zhu}},\ }\bibfield  {title} {\bibinfo {title} {Theory of {Trotter} {Error} with {Commutator} {Scaling}},\ }\href {https://doi.org/10.1103/PhysRevX.11.011020} {\bibfield  {journal} {\bibinfo  {journal} {Physical Review X}\ }\textbf {\bibinfo {volume} {11}},\ \bibinfo {pages} {011020} (\bibinfo {year} {2021}{\natexlab{b}})},\ \bibinfo {note} {publisher: American Physical Society}\BibitemShut {NoStop}%
\bibitem [{\citenamefont {Sawaya}\ \emph {et~al.}(2020)\citenamefont {Sawaya}, \citenamefont {Menke}, \citenamefont {Kyaw}, \citenamefont {Johri}, \citenamefont {Aspuru-Guzik},\ and\ \citenamefont {Guerreschi}}]{sawaya_resource-efficient_2020}%
  \BibitemOpen
  \bibfield  {author} {\bibinfo {author} {\bibfnamefont {N.~P.~D.}\ \bibnamefont {Sawaya}}, \bibinfo {author} {\bibfnamefont {T.}~\bibnamefont {Menke}}, \bibinfo {author} {\bibfnamefont {T.~H.}\ \bibnamefont {Kyaw}}, \bibinfo {author} {\bibfnamefont {S.}~\bibnamefont {Johri}}, \bibinfo {author} {\bibfnamefont {A.}~\bibnamefont {Aspuru-Guzik}},\ and\ \bibinfo {author} {\bibfnamefont {G.~G.}\ \bibnamefont {Guerreschi}},\ }\bibfield  {title} {\bibinfo {title} {Resource-efficient digital quantum simulation of d-level systems for photonic, vibrational, and spin-s {Hamiltonians}},\ }\href {https://doi.org/10.1038/s41534-020-0278-0} {\bibfield  {journal} {\bibinfo  {journal} {npj Quantum Information}\ }\textbf {\bibinfo {volume} {6}},\ \bibinfo {pages} {1} (\bibinfo {year} {2020})},\ \bibinfo {note} {publisher: Nature Publishing Group}\BibitemShut {NoStop}%
\bibitem [{\citenamefont {Choi}(1975)}]{choi_completely_1975}%
  \BibitemOpen
  \bibfield  {author} {\bibinfo {author} {\bibfnamefont {M.-D.}\ \bibnamefont {Choi}},\ }\bibfield  {title} {\bibinfo {title} {Completely positive linear maps on complex matrices},\ }\href {https://doi.org/10.1016/0024-3795(75)90075-0} {\bibfield  {journal} {\bibinfo  {journal} {Linear Algebra and its Applications}\ }\textbf {\bibinfo {volume} {10}},\ \bibinfo {pages} {285} (\bibinfo {year} {1975})}\BibitemShut {NoStop}%
\bibitem [{\citenamefont {Havel}(2003)}]{havel_robust_2003}%
  \BibitemOpen
  \bibfield  {author} {\bibinfo {author} {\bibfnamefont {T.~F.}\ \bibnamefont {Havel}},\ }\bibfield  {title} {\bibinfo {title} {Robust procedures for converting among {Lindblad}, {Kraus} and matrix representations of quantum dynamical semigroups},\ }\href {https://doi.org/10.1063/1.1518555} {\bibfield  {journal} {\bibinfo  {journal} {Journal of Mathematical Physics}\ }\textbf {\bibinfo {volume} {44}},\ \bibinfo {pages} {534} (\bibinfo {year} {2003})}\BibitemShut {NoStop}%
\bibitem [{\citenamefont {Burdine}\ and\ \citenamefont {Blair}(2024)}]{burdine_trotterless_2024}%
  \BibitemOpen
  \bibfield  {author} {\bibinfo {author} {\bibfnamefont {C.}~\bibnamefont {Burdine}}\ and\ \bibinfo {author} {\bibfnamefont {E.~P.}\ \bibnamefont {Blair}},\ }\bibfield  {title} {\bibinfo {title} {Trotterless {Simulation} of {Open} {Quantum} {Systems} for {NISQ} {Quantum} {Devices}},\ }\href {https://doi.org/10.1002/qute.202400240} {\bibfield  {journal} {\bibinfo  {journal} {Advanced Quantum Technologies}\ ,\ \bibinfo {pages} {2400240}} (\bibinfo {year} {2024})}\BibitemShut {NoStop}%
\bibitem [{\citenamefont {Camps}\ and\ \citenamefont {Van~Beeumen}(2022)}]{camps_fable_2022}%
  \BibitemOpen
  \bibfield  {author} {\bibinfo {author} {\bibfnamefont {D.}~\bibnamefont {Camps}}\ and\ \bibinfo {author} {\bibfnamefont {R.}~\bibnamefont {Van~Beeumen}},\ }\bibfield  {title} {\bibinfo {title} {{FABLE}: {Fast} {Approximate} {Quantum} {Circuits} for {Block}-{Encodings}},\ }in\ \href {https://doi.org/10.1109/QCE53715.2022.00029} {\emph {\bibinfo {booktitle} {2022 {IEEE} {International} {Conference} on {Quantum} {Computing} and {Engineering} ({QCE})}}}\ (\bibinfo {year} {2022})\ pp.\ \bibinfo {pages} {104--113}\BibitemShut {NoStop}%
\bibitem [{\citenamefont {Schäffer}(1955)}]{schaffer_unitary_1955}%
  \BibitemOpen
  \bibfield  {author} {\bibinfo {author} {\bibfnamefont {J.~J.}\ \bibnamefont {Schäffer}},\ }\bibfield  {title} {\bibinfo {title} {On {Unitary} {Dilations} of {Contractions}},\ }\href {https://doi.org/10.2307/2032368} {\bibfield  {journal} {\bibinfo  {journal} {Proceedings of the American Mathematical Society}\ }\textbf {\bibinfo {volume} {6}},\ \bibinfo {pages} {322} (\bibinfo {year} {1955})},\ \bibinfo {note} {publisher: American Mathematical Society}\BibitemShut {NoStop}%
\bibitem [{\citenamefont {Zheng}(2021)}]{zheng_universal_2021}%
  \BibitemOpen
  \bibfield  {author} {\bibinfo {author} {\bibfnamefont {C.}~\bibnamefont {Zheng}},\ }\bibfield  {title} {\bibinfo {title} {Universal quantum simulation of single-qubit nonunitary operators using duality quantum algorithm},\ }\href {https://doi.org/10.1038/s41598-021-83521-5} {\bibfield  {journal} {\bibinfo  {journal} {Scientific Reports}\ }\textbf {\bibinfo {volume} {11}},\ \bibinfo {pages} {3960} (\bibinfo {year} {2021})},\ \bibinfo {note} {publisher: Nature Publishing Group}\BibitemShut {NoStop}%
\bibitem [{\citenamefont {Childs}\ and\ \citenamefont {Wiebe}(2012)}]{childs_hamiltonian_2012}%
  \BibitemOpen
  \bibfield  {author} {\bibinfo {author} {\bibfnamefont {A.~M.}\ \bibnamefont {Childs}}\ and\ \bibinfo {author} {\bibfnamefont {N.}~\bibnamefont {Wiebe}},\ }\bibfield  {title} {\bibinfo {title} {Hamiltonian {Simulation} {Using} {Linear} {Combinations} of {Unitary} {Operations}}\ }\href {https://doi.org/10.48550/ARXIV.1202.5822} {10.48550/ARXIV.1202.5822} (\bibinfo {year} {2012}),\ \bibinfo {note} {publisher: arXiv Version Number: 1}\BibitemShut {NoStop}%
\bibitem [{\citenamefont {Shende}\ \emph {et~al.}(2006)\citenamefont {Shende}, \citenamefont {Bullock},\ and\ \citenamefont {Markov}}]{shende_synthesis_2006}%
  \BibitemOpen
  \bibfield  {author} {\bibinfo {author} {\bibfnamefont {V.~V.}\ \bibnamefont {Shende}}, \bibinfo {author} {\bibfnamefont {S.~S.}\ \bibnamefont {Bullock}},\ and\ \bibinfo {author} {\bibfnamefont {I.~L.}\ \bibnamefont {Markov}},\ }\bibfield  {title} {\bibinfo {title} {Synthesis of {Quantum} {Logic} {Circuits}},\ }\href {https://doi.org/10.1109/TCAD.2005.855930} {\bibfield  {journal} {\bibinfo  {journal} {IEEE Transactions on Computer-Aided Design of Integrated Circuits and Systems}\ }\textbf {\bibinfo {volume} {25}},\ \bibinfo {pages} {1000} (\bibinfo {year} {2006})},\ \bibinfo {note} {arXiv:quant-ph/0406176}\BibitemShut {NoStop}%
\bibitem [{\citenamefont {Cai}\ \emph {et~al.}(2016)\citenamefont {Cai}, \citenamefont {Kim}, \citenamefont {Wang}, \citenamefont {Yuan},\ and\ \citenamefont {Zhou}}]{cai_optimal_2016}%
  \BibitemOpen
  \bibfield  {author} {\bibinfo {author} {\bibfnamefont {T.}~\bibnamefont {Cai}}, \bibinfo {author} {\bibfnamefont {D.}~\bibnamefont {Kim}}, \bibinfo {author} {\bibfnamefont {Y.}~\bibnamefont {Wang}}, \bibinfo {author} {\bibfnamefont {M.}~\bibnamefont {Yuan}},\ and\ \bibinfo {author} {\bibfnamefont {H.~H.}\ \bibnamefont {Zhou}},\ }\bibfield  {title} {\bibinfo {title} {Optimal large-scale quantum state tomography with {Pauli} measurements},\ }\href {https://doi.org/10.1214/15-AOS1382} {\bibfield  {journal} {\bibinfo  {journal} {The Annals of Statistics}\ }\textbf {\bibinfo {volume} {44}},\ \bibinfo {pages} {682} (\bibinfo {year} {2016})},\ \bibinfo {note} {publisher: Institute of Mathematical Statistics}\BibitemShut {NoStop}%
\bibitem [{\citenamefont {Schwinger}(1952)}]{schwinger_angular_1952}%
  \BibitemOpen
  \bibfield  {author} {\bibinfo {author} {\bibfnamefont {J.}~\bibnamefont {Schwinger}},\ }\href {https://doi.org/10.2172/4389568} {\emph {\bibinfo {title} {On {Angular} {Momentum}}}},\ \bibinfo {type} {Tech. Rep.}\ \bibinfo {number} {NYO-3071}\ (\bibinfo  {institution} {Harvard Univ., Cambridge, MA (United States); Nuclear Development Associates, Inc. (US)},\ \bibinfo {year} {1952})\BibitemShut {NoStop}%
\bibitem [{\citenamefont {Fabre}\ and\ \citenamefont {Treps}(2020)}]{fabre_modes_2020}%
  \BibitemOpen
  \bibfield  {author} {\bibinfo {author} {\bibfnamefont {C.}~\bibnamefont {Fabre}}\ and\ \bibinfo {author} {\bibfnamefont {N.}~\bibnamefont {Treps}},\ }\bibfield  {title} {\bibinfo {title} {Modes and states in quantum optics},\ }\href {https://doi.org/10.1103/RevModPhys.92.035005} {\bibfield  {journal} {\bibinfo  {journal} {Reviews of Modern Physics}\ }\textbf {\bibinfo {volume} {92}},\ \bibinfo {pages} {035005} (\bibinfo {year} {2020})},\ \bibinfo {note} {publisher: American Physical Society}\BibitemShut {NoStop}%
\bibitem [{\citenamefont {Bolmatov}\ \emph {et~al.}(2015)\citenamefont {Bolmatov}, \citenamefont {Zav’yalov}, \citenamefont {Zhernenkov}, \citenamefont {Musaev},\ and\ \citenamefont {Cai}}]{bolmatov_unified_2015}%
  \BibitemOpen
  \bibfield  {author} {\bibinfo {author} {\bibfnamefont {D.}~\bibnamefont {Bolmatov}}, \bibinfo {author} {\bibfnamefont {D.}~\bibnamefont {Zav’yalov}}, \bibinfo {author} {\bibfnamefont {M.}~\bibnamefont {Zhernenkov}}, \bibinfo {author} {\bibfnamefont {E.~T.}\ \bibnamefont {Musaev}},\ and\ \bibinfo {author} {\bibfnamefont {Y.~Q.}\ \bibnamefont {Cai}},\ }\bibfield  {title} {\bibinfo {title} {Unified phonon-based approach to the thermodynamics of solid, liquid and gas states},\ }\href {https://doi.org/10.1016/j.aop.2015.09.018} {\bibfield  {journal} {\bibinfo  {journal} {Annals of Physics}\ }\textbf {\bibinfo {volume} {363}},\ \bibinfo {pages} {221} (\bibinfo {year} {2015})}\BibitemShut {NoStop}%
\bibitem [{\citenamefont {Xiao}(2009)}]{xiao_theory_2009}%
  \BibitemOpen
  \bibfield  {author} {\bibinfo {author} {\bibfnamefont {M.-w.}\ \bibnamefont {Xiao}},\ }\href {https://doi.org/10.48550/arXiv.0908.0787} {\bibinfo {title} {Theory of transformation for the diagonalization of quadratic {Hamiltonians}}} (\bibinfo {year} {2009}),\ \bibinfo {note} {arXiv:0908.0787 [math-ph]}\BibitemShut {NoStop}%
\bibitem [{\citenamefont {Vendromin}\ and\ \citenamefont {Dignam}(2022)}]{Vendromin_2022}%
  \BibitemOpen
  \bibfield  {author} {\bibinfo {author} {\bibfnamefont {C.}~\bibnamefont {Vendromin}}\ and\ \bibinfo {author} {\bibfnamefont {M.~M.}\ \bibnamefont {Dignam}},\ }\bibfield  {title} {\bibinfo {title} {Simple way to incorporate loss when modeling multimode-entangled-state generation},\ }\href {https://doi.org/10.1103/PhysRevA.105.063707} {\bibfield  {journal} {\bibinfo  {journal} {Phys. Rev. A}\ }\textbf {\bibinfo {volume} {105}},\ \bibinfo {pages} {063707} (\bibinfo {year} {2022})}\BibitemShut {NoStop}%
\bibitem [{\citenamefont {Seifoory}\ \emph {et~al.}(2017)\citenamefont {Seifoory}, \citenamefont {Doutre}, \citenamefont {Dignam},\ and\ \citenamefont {Sipe}}]{seifoory_2017}%
  \BibitemOpen
  \bibfield  {author} {\bibinfo {author} {\bibfnamefont {H.}~\bibnamefont {Seifoory}}, \bibinfo {author} {\bibfnamefont {S.}~\bibnamefont {Doutre}}, \bibinfo {author} {\bibfnamefont {M.}~\bibnamefont {Dignam}},\ and\ \bibinfo {author} {\bibfnamefont {J.}~\bibnamefont {Sipe}},\ }\bibfield  {title} {\bibinfo {title} {Squeezed thermal states: the result of parametric down conversion in lossy cavities},\ }\href {https://doi.org/10.1364/JOSAB.34.001587} {\bibfield  {journal} {\bibinfo  {journal} {Journal of the Optical Society of America B}\ }\textbf {\bibinfo {volume} {34}},\ \bibinfo {pages} {1587} (\bibinfo {year} {2017})}\BibitemShut {NoStop}%
\bibitem [{\citenamefont {Parto}\ \emph {et~al.}(2023)\citenamefont {Parto}, \citenamefont {Leefmans}, \citenamefont {Williams}, \citenamefont {Nori},\ and\ \citenamefont {Marandi}}]{Parto_2023}%
  \BibitemOpen
  \bibfield  {author} {\bibinfo {author} {\bibfnamefont {M.}~\bibnamefont {Parto}}, \bibinfo {author} {\bibfnamefont {C.}~\bibnamefont {Leefmans}}, \bibinfo {author} {\bibfnamefont {J.}~\bibnamefont {Williams}}, \bibinfo {author} {\bibfnamefont {F.}~\bibnamefont {Nori}},\ and\ \bibinfo {author} {\bibfnamefont {A.}~\bibnamefont {Marandi}},\ }\bibfield  {title} {\bibinfo {title} {Non-abelian effects in dissipative photonic topological lattices},\ }\bibfield  {journal} {\bibinfo  {journal} {Nature Communications}\ }\textbf {\bibinfo {volume} {14}},\ \href {https://doi.org/10.1038/s41467-023-37065-z} {10.1038/s41467-023-37065-z} (\bibinfo {year} {2023})\BibitemShut {NoStop}%
\bibitem [{\citenamefont {Sturges}\ \emph {et~al.}(2021)\citenamefont {Sturges}, \citenamefont {McDermott}, \citenamefont {Buraczewski}, \citenamefont {Clements}, \citenamefont {Renema}, \citenamefont {Nam}, \citenamefont {Gerrits}, \citenamefont {Lita}, \citenamefont {Kolthammer}, \citenamefont {Eckstein}, \citenamefont {Walmsley},\ and\ \citenamefont {Stobińska}}]{Sturges_2021}%
  \BibitemOpen
  \bibfield  {author} {\bibinfo {author} {\bibfnamefont {T.~J.}\ \bibnamefont {Sturges}}, \bibinfo {author} {\bibfnamefont {T.}~\bibnamefont {McDermott}}, \bibinfo {author} {\bibfnamefont {A.}~\bibnamefont {Buraczewski}}, \bibinfo {author} {\bibfnamefont {W.~R.}\ \bibnamefont {Clements}}, \bibinfo {author} {\bibfnamefont {J.~J.}\ \bibnamefont {Renema}}, \bibinfo {author} {\bibfnamefont {S.~W.}\ \bibnamefont {Nam}}, \bibinfo {author} {\bibfnamefont {T.}~\bibnamefont {Gerrits}}, \bibinfo {author} {\bibfnamefont {A.}~\bibnamefont {Lita}}, \bibinfo {author} {\bibfnamefont {W.~S.}\ \bibnamefont {Kolthammer}}, \bibinfo {author} {\bibfnamefont {A.}~\bibnamefont {Eckstein}}, \bibinfo {author} {\bibfnamefont {I.~A.}\ \bibnamefont {Walmsley}},\ and\ \bibinfo {author} {\bibfnamefont {M.}~\bibnamefont {Stobińska}},\ }\bibfield  {title} {\bibinfo {title} {Quantum simulations with multiphoton fock states},\ }\bibfield  {journal} {\bibinfo  {journal} {npj Quantum Information}\ }\textbf {\bibinfo {volume} {7}},\ \href
  {https://doi.org/10.1038/s41534-021-00427-w} {10.1038/s41534-021-00427-w} (\bibinfo {year} {2021})\BibitemShut {NoStop}%
\bibitem [{\citenamefont {Araujo}\ \emph {et~al.}(2021)\citenamefont {Araujo}, \citenamefont {Park}, \citenamefont {Petruccione},\ and\ \citenamefont {da~Silva}}]{araujo_divide-and-conquer_2021}%
  \BibitemOpen
  \bibfield  {author} {\bibinfo {author} {\bibfnamefont {I.~F.}\ \bibnamefont {Araujo}}, \bibinfo {author} {\bibfnamefont {D.~K.}\ \bibnamefont {Park}}, \bibinfo {author} {\bibfnamefont {F.}~\bibnamefont {Petruccione}},\ and\ \bibinfo {author} {\bibfnamefont {A.~J.}\ \bibnamefont {da~Silva}},\ }\bibfield  {title} {\bibinfo {title} {A divide-and-conquer algorithm for quantum state preparation},\ }\href {https://doi.org/10.1038/s41598-021-85474-1} {\bibfield  {journal} {\bibinfo  {journal} {Scientific Reports}\ }\textbf {\bibinfo {volume} {11}},\ \bibinfo {pages} {6329} (\bibinfo {year} {2021})},\ \bibinfo {note} {publisher: Nature Publishing Group}\BibitemShut {NoStop}%
\bibitem [{noa(2024{\natexlab{a}})}]{noauthor_ionq_2024}%
  \BibitemOpen
  \href {https://ionq.com/quantum-systems/harmony} {\bibinfo {title} {{IonQ} {Harmony}}} (\bibinfo {year} {2024}{\natexlab{a}})\BibitemShut {NoStop}%
\bibitem [{noa(2024{\natexlab{b}})}]{noauthor_quantinuum_2024}%
  \BibitemOpen
  \href {https://quantinuum.com/} {\bibinfo {title} {Quantinuum {H1}-1}} (\bibinfo {year} {2024}{\natexlab{b}})\BibitemShut {NoStop}%
\bibitem [{\citenamefont {Wright}\ \emph {et~al.}(2019)\citenamefont {Wright}, \citenamefont {Beck}, \citenamefont {Debnath}, \citenamefont {Amini}, \citenamefont {Nam}, \citenamefont {Grzesiak}, \citenamefont {Chen}, \citenamefont {Pisenti}, \citenamefont {Chmielewski}, \citenamefont {Collins}, \citenamefont {Hudek}, \citenamefont {Mizrahi}, \citenamefont {Wong-Campos}, \citenamefont {Allen}, \citenamefont {Apisdorf}, \citenamefont {Solomon}, \citenamefont {Williams}, \citenamefont {Ducore}, \citenamefont {Blinov}, \citenamefont {Kreikemeier}, \citenamefont {Chaplin}, \citenamefont {Keesan}, \citenamefont {Monroe},\ and\ \citenamefont {Kim}}]{wright_benchmarking_2019}%
  \BibitemOpen
  \bibfield  {author} {\bibinfo {author} {\bibfnamefont {K.}~\bibnamefont {Wright}}, \bibinfo {author} {\bibfnamefont {K.~M.}\ \bibnamefont {Beck}}, \bibinfo {author} {\bibfnamefont {S.}~\bibnamefont {Debnath}}, \bibinfo {author} {\bibfnamefont {J.~M.}\ \bibnamefont {Amini}}, \bibinfo {author} {\bibfnamefont {Y.}~\bibnamefont {Nam}}, \bibinfo {author} {\bibfnamefont {N.}~\bibnamefont {Grzesiak}}, \bibinfo {author} {\bibfnamefont {J.-S.}\ \bibnamefont {Chen}}, \bibinfo {author} {\bibfnamefont {N.~C.}\ \bibnamefont {Pisenti}}, \bibinfo {author} {\bibfnamefont {M.}~\bibnamefont {Chmielewski}}, \bibinfo {author} {\bibfnamefont {C.}~\bibnamefont {Collins}}, \bibinfo {author} {\bibfnamefont {K.~M.}\ \bibnamefont {Hudek}}, \bibinfo {author} {\bibfnamefont {J.}~\bibnamefont {Mizrahi}}, \bibinfo {author} {\bibfnamefont {J.~D.}\ \bibnamefont {Wong-Campos}}, \bibinfo {author} {\bibfnamefont {S.}~\bibnamefont {Allen}}, \bibinfo {author} {\bibfnamefont {J.}~\bibnamefont {Apisdorf}}, \bibinfo {author} {\bibfnamefont
  {P.}~\bibnamefont {Solomon}}, \bibinfo {author} {\bibfnamefont {M.}~\bibnamefont {Williams}}, \bibinfo {author} {\bibfnamefont {A.~M.}\ \bibnamefont {Ducore}}, \bibinfo {author} {\bibfnamefont {A.}~\bibnamefont {Blinov}}, \bibinfo {author} {\bibfnamefont {S.~M.}\ \bibnamefont {Kreikemeier}}, \bibinfo {author} {\bibfnamefont {V.}~\bibnamefont {Chaplin}}, \bibinfo {author} {\bibfnamefont {M.}~\bibnamefont {Keesan}}, \bibinfo {author} {\bibfnamefont {C.}~\bibnamefont {Monroe}},\ and\ \bibinfo {author} {\bibfnamefont {J.}~\bibnamefont {Kim}},\ }\bibfield  {title} {\bibinfo {title} {Benchmarking an 11-qubit quantum computer},\ }\href {https://doi.org/10.1038/s41467-019-13534-2} {\bibfield  {journal} {\bibinfo  {journal} {Nature Communications}\ }\textbf {\bibinfo {volume} {10}},\ \bibinfo {pages} {5464} (\bibinfo {year} {2019})},\ \bibinfo {note} {publisher: Nature Publishing Group}\BibitemShut {NoStop}%
\bibitem [{\citenamefont {Javadi-Abhari}\ \emph {et~al.}(2024)\citenamefont {Javadi-Abhari}, \citenamefont {Treinish}, \citenamefont {Krsulich}, \citenamefont {Wood}, \citenamefont {Lishman}, \citenamefont {Gacon}, \citenamefont {Martiel}, \citenamefont {Nation}, \citenamefont {Bishop}, \citenamefont {Cross}, \citenamefont {Johnson},\ and\ \citenamefont {Gambetta}}]{javadi-abhari_quantum_2024}%
  \BibitemOpen
  \bibfield  {author} {\bibinfo {author} {\bibfnamefont {A.}~\bibnamefont {Javadi-Abhari}}, \bibinfo {author} {\bibfnamefont {M.}~\bibnamefont {Treinish}}, \bibinfo {author} {\bibfnamefont {K.}~\bibnamefont {Krsulich}}, \bibinfo {author} {\bibfnamefont {C.~J.}\ \bibnamefont {Wood}}, \bibinfo {author} {\bibfnamefont {J.}~\bibnamefont {Lishman}}, \bibinfo {author} {\bibfnamefont {J.}~\bibnamefont {Gacon}}, \bibinfo {author} {\bibfnamefont {S.}~\bibnamefont {Martiel}}, \bibinfo {author} {\bibfnamefont {P.~D.}\ \bibnamefont {Nation}}, \bibinfo {author} {\bibfnamefont {L.~S.}\ \bibnamefont {Bishop}}, \bibinfo {author} {\bibfnamefont {A.~W.}\ \bibnamefont {Cross}}, \bibinfo {author} {\bibfnamefont {B.~R.}\ \bibnamefont {Johnson}},\ and\ \bibinfo {author} {\bibfnamefont {J.~M.}\ \bibnamefont {Gambetta}},\ }\href {https://doi.org/10.48550/arXiv.2405.08810} {\bibinfo {title} {Quantum computing with {Qiskit}}} (\bibinfo {year} {2024}),\ \bibinfo {note} {arXiv:2405.08810 [quant-ph]}\BibitemShut {NoStop}%
\bibitem [{\citenamefont {Sivarajah}\ \emph {et~al.}(2020)\citenamefont {Sivarajah}, \citenamefont {Dilkes}, \citenamefont {Cowtan}, \citenamefont {Simmons}, \citenamefont {Edgington},\ and\ \citenamefont {Duncan}}]{Sivarajah_2021}%
  \BibitemOpen
  \bibfield  {author} {\bibinfo {author} {\bibfnamefont {S.}~\bibnamefont {Sivarajah}}, \bibinfo {author} {\bibfnamefont {S.}~\bibnamefont {Dilkes}}, \bibinfo {author} {\bibfnamefont {A.}~\bibnamefont {Cowtan}}, \bibinfo {author} {\bibfnamefont {W.}~\bibnamefont {Simmons}}, \bibinfo {author} {\bibfnamefont {A.}~\bibnamefont {Edgington}},\ and\ \bibinfo {author} {\bibfnamefont {R.}~\bibnamefont {Duncan}},\ }\bibfield  {title} {\bibinfo {title} {t|ket⟩: a retargetable compiler for nisq devices},\ }\href {https://doi.org/10.1088/2058-9565/ab8e92} {\bibfield  {journal} {\bibinfo  {journal} {Quantum Science and Technology}\ }\textbf {\bibinfo {volume} {6}},\ \bibinfo {pages} {014003} (\bibinfo {year} {2020})}\BibitemShut {NoStop}%
\bibitem [{\citenamefont {David}\ \emph {et~al.}(2024)\citenamefont {David}, \citenamefont {Sinayskiy},\ and\ \citenamefont {Petruccione}}]{david_digital_2024}%
  \BibitemOpen
  \bibfield  {author} {\bibinfo {author} {\bibfnamefont {I.~J.}\ \bibnamefont {David}}, \bibinfo {author} {\bibfnamefont {I.}~\bibnamefont {Sinayskiy}},\ and\ \bibinfo {author} {\bibfnamefont {F.}~\bibnamefont {Petruccione}},\ }\bibfield  {title} {\bibinfo {title} {Digital simulation of convex mixtures of {Markovian} and non-{Markovian} single qubit {Pauli} channels on {NISQ} devices},\ }\href {https://doi.org/10.1140/epjqt/s40507-024-00224-2} {\bibfield  {journal} {\bibinfo  {journal} {EPJ Quantum Technology}\ }\textbf {\bibinfo {volume} {11}},\ \bibinfo {pages} {1} (\bibinfo {year} {2024})},\ \bibinfo {note} {number: 1 Publisher: SpringerOpen}\BibitemShut {NoStop}%
\bibitem [{\citenamefont {Bhattacharya}\ \emph {et~al.}(2017)\citenamefont {Bhattacharya}, \citenamefont {Misra}, \citenamefont {Mukhopadhyay},\ and\ \citenamefont {Pati}}]{bhattacharya_exact_2017}%
  \BibitemOpen
  \bibfield  {author} {\bibinfo {author} {\bibfnamefont {S.}~\bibnamefont {Bhattacharya}}, \bibinfo {author} {\bibfnamefont {A.}~\bibnamefont {Misra}}, \bibinfo {author} {\bibfnamefont {C.}~\bibnamefont {Mukhopadhyay}},\ and\ \bibinfo {author} {\bibfnamefont {A.~K.}\ \bibnamefont {Pati}},\ }\bibfield  {title} {\bibinfo {title} {Exact master equation for a spin interacting with a spin bath: {Non}-{Markovianity} and negative entropy production rate},\ }\href {https://doi.org/10.1103/PhysRevA.95.012122} {\bibfield  {journal} {\bibinfo  {journal} {Physical Review A}\ }\textbf {\bibinfo {volume} {95}},\ \bibinfo {pages} {012122} (\bibinfo {year} {2017})},\ \bibinfo {note} {publisher: American Physical Society}\BibitemShut {NoStop}%
\bibitem [{\citenamefont {Laikhtman}\ and\ \citenamefont {Altshuler}(1994)}]{laikhtman_quasiclassical_1994}%
  \BibitemOpen
  \bibfield  {author} {\bibinfo {author} {\bibfnamefont {B.}~\bibnamefont {Laikhtman}}\ and\ \bibinfo {author} {\bibfnamefont {E.~L.}\ \bibnamefont {Altshuler}},\ }\bibfield  {title} {\bibinfo {title} {Quasiclassical {Theory} of {Shubnikov}-de {Haas} {Effect} in {2D} {Electron} {Gas}},\ }\href {https://doi.org/10.1006/aphy.1994.1056} {\bibfield  {journal} {\bibinfo  {journal} {Annals of Physics}\ }\textbf {\bibinfo {volume} {232}},\ \bibinfo {pages} {332} (\bibinfo {year} {1994})}\BibitemShut {NoStop}%
\bibitem [{\citenamefont {Hatsugai}(1993)}]{hatsugai_chern_1993}%
  \BibitemOpen
  \bibfield  {author} {\bibinfo {author} {\bibfnamefont {Y.}~\bibnamefont {Hatsugai}},\ }\bibfield  {title} {\bibinfo {title} {Chern number and edge states in the integer quantum {Hall} effect},\ }\href {https://doi.org/10.1103/PhysRevLett.71.3697} {\bibfield  {journal} {\bibinfo  {journal} {Physical Review Letters}\ }\textbf {\bibinfo {volume} {71}},\ \bibinfo {pages} {3697} (\bibinfo {year} {1993})},\ \bibinfo {note} {publisher: American Physical Society}\BibitemShut {NoStop}%
\bibitem [{\citenamefont {Sun}\ \emph {et~al.}(2021)\citenamefont {Sun}, \citenamefont {Yuan}, \citenamefont {Tsunoda}, \citenamefont {Vedral}, \citenamefont {Benjamin},\ and\ \citenamefont {Endo}}]{sun_mitigating_2021}%
  \BibitemOpen
  \bibfield  {author} {\bibinfo {author} {\bibfnamefont {J.}~\bibnamefont {Sun}}, \bibinfo {author} {\bibfnamefont {X.}~\bibnamefont {Yuan}}, \bibinfo {author} {\bibfnamefont {T.}~\bibnamefont {Tsunoda}}, \bibinfo {author} {\bibfnamefont {V.}~\bibnamefont {Vedral}}, \bibinfo {author} {\bibfnamefont {S.~C.}\ \bibnamefont {Benjamin}},\ and\ \bibinfo {author} {\bibfnamefont {S.}~\bibnamefont {Endo}},\ }\bibfield  {title} {\bibinfo {title} {Mitigating {Realistic} {Noise} in {Practical} {Noisy} {Intermediate}-{Scale} {Quantum} {Devices}},\ }\href {https://doi.org/10.1103/PhysRevApplied.15.034026} {\bibfield  {journal} {\bibinfo  {journal} {Physical Review Applied}\ }\textbf {\bibinfo {volume} {15}},\ \bibinfo {pages} {034026} (\bibinfo {year} {2021})},\ \bibinfo {note} {publisher: American Physical Society}\BibitemShut {NoStop}%
\bibitem [{\citenamefont {Cai}\ \emph {et~al.}(2023)\citenamefont {Cai}, \citenamefont {Babbush}, \citenamefont {Benjamin}, \citenamefont {Endo}, \citenamefont {Huggins}, \citenamefont {Li}, \citenamefont {McClean},\ and\ \citenamefont {O’Brien}}]{cai_quantum_2023}%
  \BibitemOpen
  \bibfield  {author} {\bibinfo {author} {\bibfnamefont {Z.}~\bibnamefont {Cai}}, \bibinfo {author} {\bibfnamefont {R.}~\bibnamefont {Babbush}}, \bibinfo {author} {\bibfnamefont {S.~C.}\ \bibnamefont {Benjamin}}, \bibinfo {author} {\bibfnamefont {S.}~\bibnamefont {Endo}}, \bibinfo {author} {\bibfnamefont {W.~J.}\ \bibnamefont {Huggins}}, \bibinfo {author} {\bibfnamefont {Y.}~\bibnamefont {Li}}, \bibinfo {author} {\bibfnamefont {J.~R.}\ \bibnamefont {McClean}},\ and\ \bibinfo {author} {\bibfnamefont {T.~E.}\ \bibnamefont {O’Brien}},\ }\bibfield  {title} {\bibinfo {title} {Quantum error mitigation},\ }\href {https://doi.org/10.1103/RevModPhys.95.045005} {\bibfield  {journal} {\bibinfo  {journal} {Reviews of Modern Physics}\ }\textbf {\bibinfo {volume} {95}},\ \bibinfo {pages} {045005} (\bibinfo {year} {2023})},\ \bibinfo {note} {publisher: American Physical Society}\BibitemShut {NoStop}%
\bibitem [{\citenamefont {Mandilara}\ \emph {et~al.}(2009)\citenamefont {Mandilara}, \citenamefont {Karpov},\ and\ \citenamefont {Cerf}}]{mandilara_extending_2009}%
  \BibitemOpen
  \bibfield  {author} {\bibinfo {author} {\bibfnamefont {A.}~\bibnamefont {Mandilara}}, \bibinfo {author} {\bibfnamefont {E.}~\bibnamefont {Karpov}},\ and\ \bibinfo {author} {\bibfnamefont {N.~J.}\ \bibnamefont {Cerf}},\ }\bibfield  {title} {\bibinfo {title} {Extending {Hudson}'s theorem to mixed quantum states},\ }\href {https://doi.org/10.1103/PhysRevA.79.062302} {\bibfield  {journal} {\bibinfo  {journal} {Physical Review A}\ }\textbf {\bibinfo {volume} {79}},\ \bibinfo {pages} {062302} (\bibinfo {year} {2009})},\ \bibinfo {note} {arXiv:0808.2501 [quant-ph]}\BibitemShut {NoStop}%
\bibitem [{\citenamefont {Bartlett}\ and\ \citenamefont {Moyal}(1949)}]{bartlett_exact_1949}%
  \BibitemOpen
  \bibfield  {author} {\bibinfo {author} {\bibfnamefont {M.~S.}\ \bibnamefont {Bartlett}}\ and\ \bibinfo {author} {\bibfnamefont {J.~E.}\ \bibnamefont {Moyal}},\ }\bibfield  {title} {\bibinfo {title} {The exact transition probabilities of quantum-mechanical oscillators calculated by the phase-space method},\ }\href {https://doi.org/10.1017/S030500410002524X} {\bibfield  {journal} {\bibinfo  {journal} {Mathematical Proceedings of the Cambridge Philosophical Society}\ }\textbf {\bibinfo {volume} {45}},\ \bibinfo {pages} {545} (\bibinfo {year} {1949})}\BibitemShut {NoStop}%
\bibitem [{\citenamefont {Cochrane}\ \emph {et~al.}(1999)\citenamefont {Cochrane}, \citenamefont {Milburn},\ and\ \citenamefont {Munro}}]{cochrane_macroscopically_1999}%
  \BibitemOpen
  \bibfield  {author} {\bibinfo {author} {\bibfnamefont {P.~T.}\ \bibnamefont {Cochrane}}, \bibinfo {author} {\bibfnamefont {G.~J.}\ \bibnamefont {Milburn}},\ and\ \bibinfo {author} {\bibfnamefont {W.~J.}\ \bibnamefont {Munro}},\ }\bibfield  {title} {\bibinfo {title} {Macroscopically distinct quantum-superposition states as a bosonic code for amplitude damping},\ }\href {https://doi.org/10.1103/PhysRevA.59.2631} {\bibfield  {journal} {\bibinfo  {journal} {Physical Review A}\ }\textbf {\bibinfo {volume} {59}},\ \bibinfo {pages} {2631} (\bibinfo {year} {1999})},\ \bibinfo {note} {publisher: American Physical Society}\BibitemShut {NoStop}%
\bibitem [{\citenamefont {Michael}\ \emph {et~al.}(2016)\citenamefont {Michael}, \citenamefont {Silveri}, \citenamefont {Brierley}, \citenamefont {Albert}, \citenamefont {Salmilehto}, \citenamefont {Jiang},\ and\ \citenamefont {Girvin}}]{michael_new_2016}%
  \BibitemOpen
  \bibfield  {author} {\bibinfo {author} {\bibfnamefont {M.~H.}\ \bibnamefont {Michael}}, \bibinfo {author} {\bibfnamefont {M.}~\bibnamefont {Silveri}}, \bibinfo {author} {\bibfnamefont {R.}~\bibnamefont {Brierley}}, \bibinfo {author} {\bibfnamefont {V.~V.}\ \bibnamefont {Albert}}, \bibinfo {author} {\bibfnamefont {J.}~\bibnamefont {Salmilehto}}, \bibinfo {author} {\bibfnamefont {L.}~\bibnamefont {Jiang}},\ and\ \bibinfo {author} {\bibfnamefont {S.}~\bibnamefont {Girvin}},\ }\bibfield  {title} {\bibinfo {title} {New {Class} of {Quantum} {Error}-{Correcting} {Codes} for a {Bosonic} {Mode}},\ }\href {https://doi.org/10.1103/PhysRevX.6.031006} {\bibfield  {journal} {\bibinfo  {journal} {Physical Review X}\ }\textbf {\bibinfo {volume} {6}},\ \bibinfo {pages} {031006} (\bibinfo {year} {2016})},\ \bibinfo {note} {publisher: American Physical Society}\BibitemShut {NoStop}%
\bibitem [{\citenamefont {Head-Marsden}\ \emph {et~al.}(2021)\citenamefont {Head-Marsden}, \citenamefont {Krastanov}, \citenamefont {Mazziotti},\ and\ \citenamefont {Narang}}]{Head-Marsden_2021}%
  \BibitemOpen
  \bibfield  {author} {\bibinfo {author} {\bibfnamefont {K.}~\bibnamefont {Head-Marsden}}, \bibinfo {author} {\bibfnamefont {S.}~\bibnamefont {Krastanov}}, \bibinfo {author} {\bibfnamefont {D.~A.}\ \bibnamefont {Mazziotti}},\ and\ \bibinfo {author} {\bibfnamefont {P.}~\bibnamefont {Narang}},\ }\bibfield  {title} {\bibinfo {title} {Capturing non-markovian dynamics on near-term quantum computers},\ }\href {https://doi.org/10.1103/PhysRevResearch.3.013182} {\bibfield  {journal} {\bibinfo  {journal} {Phys. Rev. Res.}\ }\textbf {\bibinfo {volume} {3}},\ \bibinfo {pages} {013182} (\bibinfo {year} {2021})}\BibitemShut {NoStop}%
\bibitem [{\citenamefont {Guo}\ and\ \citenamefont {Gao}(2024)}]{guo2024quantumsimulationopenquantum}%
  \BibitemOpen
  \bibfield  {author} {\bibinfo {author} {\bibfnamefont {Y.}~\bibnamefont {Guo}}\ and\ \bibinfo {author} {\bibfnamefont {X.}~\bibnamefont {Gao}},\ }\href {https://arxiv.org/abs/2404.10655} {\bibinfo {title} {Quantum simulation of open quantum dynamics via non-markovian quantum state diffusion}} (\bibinfo {year} {2024}),\ \Eprint {https://arxiv.org/abs/2404.10655} {arXiv:2404.10655 [quant-ph]} \BibitemShut {NoStop}%
\bibitem [{\citenamefont {Burdine}(2024)}]{burdine_2024}%
  \BibitemOpen
  \bibfield  {author} {\bibinfo {author} {\bibfnamefont {C.}~\bibnamefont {Burdine}},\ }\href {https://github.com/cburdine/nisq-open-quantum-systems} {}\bibinfo {howpublished} {\url{https://github.com/cburdine/nisq-open-quantum-systems}} (\bibinfo {year} {2024})\BibitemShut {NoStop}%
\bibitem [{\citenamefont {Bullock}\ and\ \citenamefont {Markov}(2003)}]{bullock_smaller_2003}%
  \BibitemOpen
  \bibfield  {author} {\bibinfo {author} {\bibfnamefont {S.~S.}\ \bibnamefont {Bullock}}\ and\ \bibinfo {author} {\bibfnamefont {I.~L.}\ \bibnamefont {Markov}},\ }\href {https://doi.org/10.48550/arXiv.quant-ph/0303039} {\bibinfo {title} {Smaller {Circuits} for {Arbitrary} n-qubit {Diagonal} {Computations}}} (\bibinfo {year} {2003}),\ \bibinfo {note} {arXiv:quant-ph/0303039}\BibitemShut {NoStop}%
\end{thebibliography}%

\end{document}